\documentclass[a4paper,11pt]{article}
\pdfoutput=1 

\usepackage{jinstpub} 
\usepackage{subfigure}
\usepackage{float}
\usepackage{color,soul}
\usepackage{url}
\usepackage{lineno}

\title{\boldmath Cryogenic R5912-20Mod Photomultiplier Tube Characterization for the ProtoDUNE Dual Phase Detector}

\author[a]{D.~Belver,}
\author[a]{ E.~Calvo,}
\author[a,1]{C.~Cuesta} \note{Corresponding author.}
\author[a]{A.~Gallego-Ros,} 
\author[a]{I.~Gil-Botella,}
\author[a]{S.~Jim\'enez,} 
\author[a]{C.~Lastoria,} 
\author[b]{T.~Lux,} 
\author[a]{C.~Palomares,} 
\author[a]{D.~Redondo,}
\author[b]{F.~Sanchez,} 
\author[a]{J.~Soto-Oton,} 
\author[a]{A.~Verdugo,}

\affiliation[a]{Centro de Investigaciones Energ\'eticas, Medioambientales y Tecnol\'ogicas (CIEMAT), Av Complutense 40, 28040, Madrid, Spain}
\affiliation[b]{F\'{i}sica d\'{}Altes Energies (IFAE) - The Barcelona Institute of Science and Technology, Campus UAB,  Bellaterra (Barcelona), Spain}

\emailAdd{clara.cuesta@ciemat.es}

\abstract{The Deep Underground Neutrino Experiment (DUNE) is a dual-site experiment for long-baseline neutrino oscillation studies, and for neutrino astrophysics and nucleon decay searches. The far detector is a 40-kton underground liquid argon time-projection-chamber (LAr TPC), in which the  photon detector system adds precise timing capabilities. The ProtoDUNE Dual-Phase detector will consist of a 6$\times$6$\times$6\,m$^3$ LAr TPC to be operated at the CERN Neutrino Platform and the photon detection system will be formed by 8-inch cryogenic photomultipliers from Hamamatsu. The PMT model (R5912-20Mod) performance at cryogenic temperature is studied including dark current, gain, and linearity with the light intensity and pulse rate. In addition, the PMT base design is validated. At cold, a decrease of the PMT amplification, or fatigue effect, is measured as the PMT output current increases, either, due to high gain, light intensity or rate. Also, the characterization results of the 40 photomultipliers to be used in ProtoDUNE Dual-Phase are presented.}


\arxivnumber{} 

\begin{document}
\maketitle
\flushbottom



\section{Introduction}
\label{sec1}

The DUNE experiment aims at addressing key questions in neutrino physics and astroparticle physics~ \cite{duneCDRv2}. It includes precision measurements of the parameters that govern neutrino oscillations with the goal of measuring the CP violating phase and the neutrino mass ordering, nucleon decay searches, and detection and measurement of the electron neutrino flux from a core-collapse supernova within our galaxy. DUNE will consist of a near detector placed at Fermilab close to the production point of the muon neutrino beam of the Long-Baseline Neutrino Facility (LBNF), and four 10\,kt fiducial mass LAr TPCs as far detector in the Sanford Underground Research Facility (SURF) at 4300\,m.w.e. depth at 1300\,km from Fermilab~\cite{duneCDRv4}. 

\begin {figure}[ht]
\includegraphics[width=0.6\textwidth]{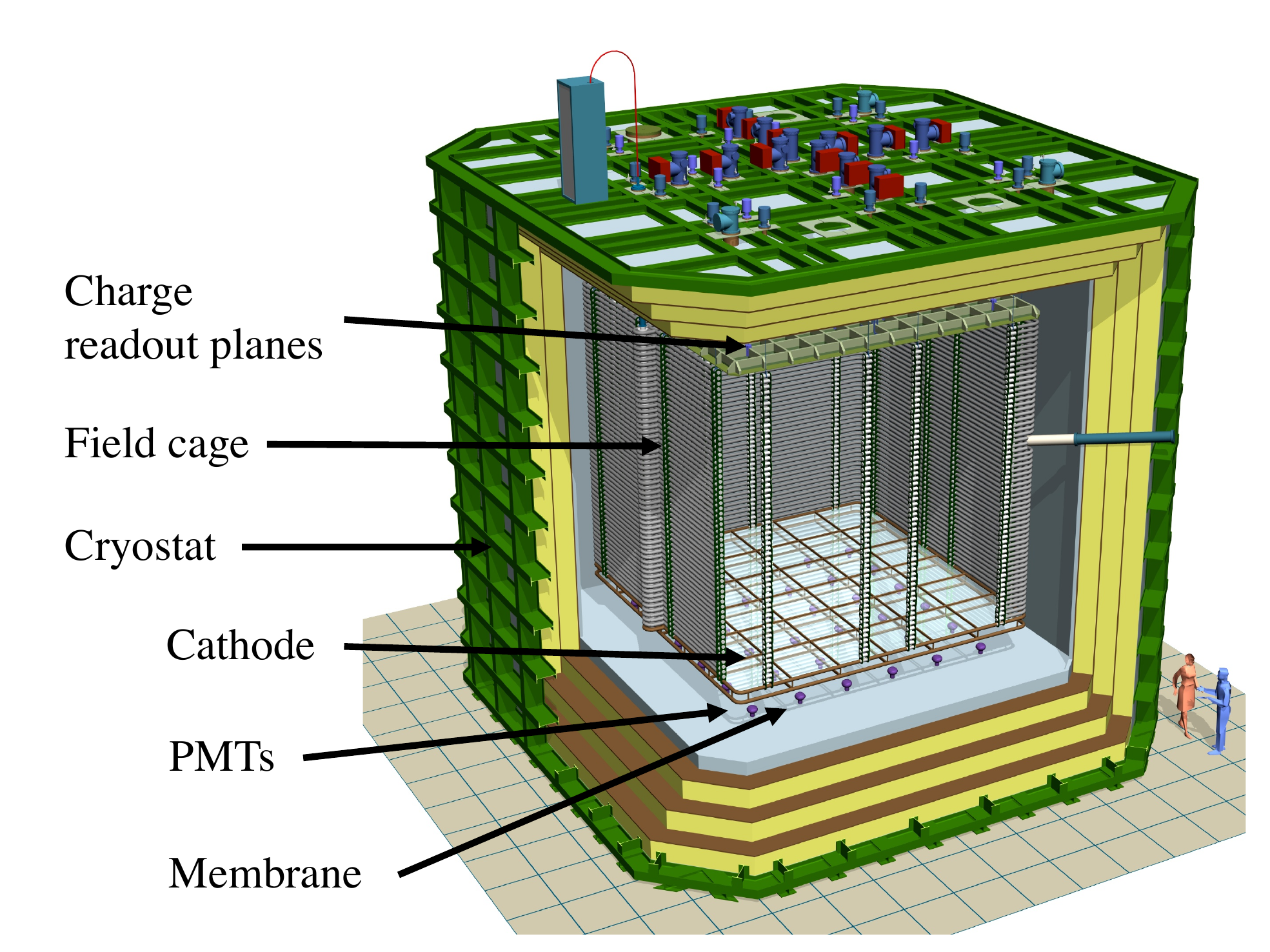}
\centering \caption{Schematic drawing of ProtoDUNE-DP where the 36 PMTs are installed at the bottom of the active volume.}
\label{fig:ProtoDUNEDP}
\end {figure}

In order to gain experience in building and operating such large-scale LAr detectors, an R\&D programme is currently underway at the CERN Neutrino Platform~\cite{ProtoDUNEs}. Such programme will operate two prototypes with the specific aim of testing the design, assembly, and installation procedures, the detector operations, as well as data acquisition, storage, processing, and analysis. The two prototypes will employ LAr TPCs as detection technology. One prototype will only use LAr, called ProtoDUNE Single-Phase, and the other will use argon in both its gaseous and liquid state, thus the name ProtoDUNE Dual-Phase (DP). Both detectors will have similar sizes. In particular, ProtoDUNE-DP~\cite{wa105}, also known as WA105 (NP02), will have an active volume of 6$\times$6$\times$6 m$^{3}$ corresponding to a mass of 300\,t. A schematic drawing of ProtoDUNE-DP is shown in Figure~\ref{fig:ProtoDUNEDP}. In ProtoDUNE-DP the charge is extracted, amplified, and detected in gaseous argon above the liquid surface allowing a finer readout pitch, a lower energy threshold, and better pattern reconstruction of the events. 

In the case of a charged particle traversing the LAr, the medium is ionized and at the same time photons are emitted. LAr scintillation light is in the far vacuum ultraviolet, having a wavelength centered at 127\,nm with a width of 8\,nm~\cite{heindl}. The scintillation light signal is used as trigger for non-beam events, to determine precisely the event time, for cosmic background rejection, and there is also a possibility to perform calorimetric measurements and particle identification. The prompt scintillation light (usually referred to as S1 signal) in LAr has two components: S1 fast of $\sim$6\,ns, and S1 slow of $\sim$1.6\,$\mu$s. In addition, the electroluminiscence, secondary scintillation light called S2, is produced in the gas phase of the detector when electrons, extracted form the liquid, are accelerated in the electric field between the liquid phase and the anode. The time length of S2 reflects the maximum drift time of original ionization
in the liquid phase up to the gas phase, thus for $\sim$1\,kV/cm electric field, the time scale of S2 is of the order of hundreds of microseconds.

The photon detection system of ProtoDUNE-DP~\cite{protoDUNElight} is formed by 36 8-inch cryogenic photomultipliers (PMTs) placed at the bottom of the active volume, submerged in LAr, behind the cathode grid. As wavelength-shifter, tetraphenyl butadiene (TPB) is coated directly on the PMTs. The photon detection system must have precision timing capabilities of few ns, a wide dynamic range to record from a few photo-electrons (p.e.) to thousands of p.e., a linear response even under high frequency or high intensity light conditions, and be able to operate at cryogenic temperature (94\,K).

This paper describes the PMT and base circuit to be used in ProtoDUNE-DP in section~\ref{sec2}. The dedicated facility used to carry out the PMT studies is presented in section~\ref{sec3}. Section~\ref{sec4} introduces the approach and procedure used in the measurements. The PMT validation measurements are detailed in section~\ref{sec5}, and the characterization results of the 36 PMTs to be used in ProtoDUNE-DP are described in section~\ref{sec6}. Finally, the results of the PMT performance at cryogenic temperatures that can be valuable for other experiments are summarized in the conclusions.


\section{PMT unit}
\label{sec2}

The photon detection system of the ProtoDUNE-DP experiment is based on Hamamatsu R5912-20Mod cryogenic photomultipliers~\cite{ham}. This 8-inch diameter PMT was selected as the light sensor since a large sensor coverage of the cathode area of 36\,m$^{2}$ should be achieved and also due to its proven reliability on other cryogenic detectors. This PMT was successfully operated in the WA105 3$\times$1$\times$1\,m$^3$ detector~\cite{311}. Also, similar PMTs were used in other LAr experiments like MicroBooNE~\cite{microboone1}, MiniCLEAN~\cite{clean2}, ArDM~\cite{ArDM2} and ICARUS T600~\cite{icarus}. 

The selected PMT has a 14-stage dynode chain which provides a nominal gain of 10$^9$ at room temperature (RT), required to compensate the gain loss at cryogenic temperatures (CT). The dynode structure is Box \& Line which is a combination of the so-called box type, which has a large collection area at the first dynode, and the linear-focused configuration, whose dynodes are designed to ensure progressive focusing of the electron paths through the multiplier. The higher number of dynodes also has the advantage of requiring a lower operation voltage for a given gain, which reduces the potential risk of sparks and the heat dissipation in the PMT bases. As the PMTs are designed to operate at CT, a thin platinum layer was added between the bi-alkali photocatode and the borosilicate glass envelope to preserve the conductivity of the photocathode at these temperatures. The cathode sensitivity provides a spectral response from 300 to 650\,nm. It is worth noting that the PMT tests described in this paper are done prior to the TPB coating.

\begin {figure}[ht!]
\includegraphics[width=0.25\textwidth]{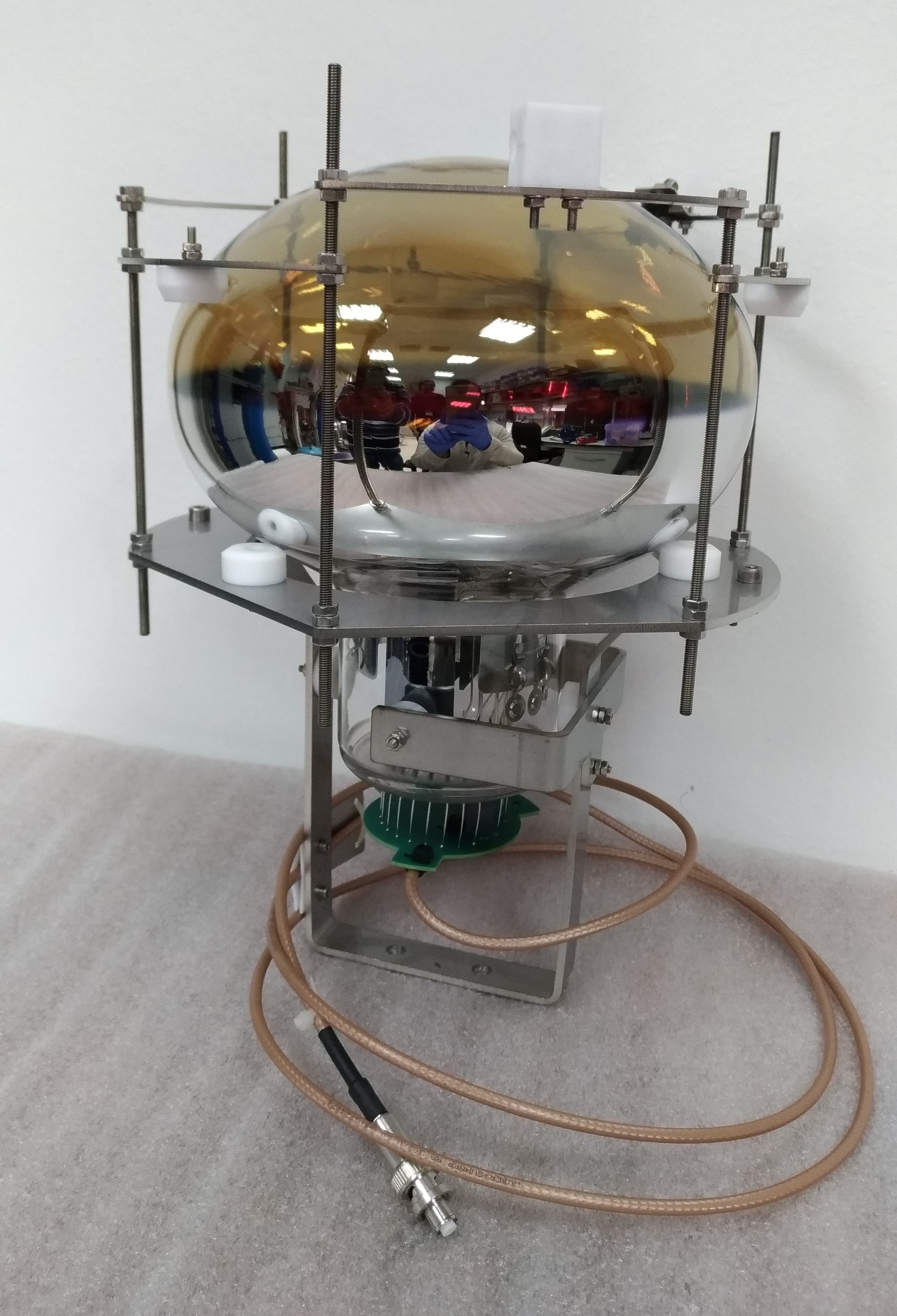}
\centering \caption{Picture of R5912-20Mod PMT with the mechanical support, the PMT base, and its cable.}
\label{fig:PMT}
\end {figure}

An individual PMT support structure was designed, manufactured and assembled at CIEMAT. This structure is mainly composed of 304L stainless steel with some small Teflon (PTFE) 6.6 pieces assembled by A4 stainless steel screws that minimize the mass while ensuring the PMT support to the membrane. The design takes into account the shrinking of the different materials during the cooling process to avoid the break of the PMT glass. The PMTs are placed at the bottom of the ProtoDUNE-DP membrane fixed to a stainless steel plate which is glued to the bottom of the cryostat. A picture of one PMT assembled on its support is shown in Figure~\ref{fig:PMT}.

\begin {figure}[ht!]
\subfigure[]{\includegraphics[width=0.75\textwidth]{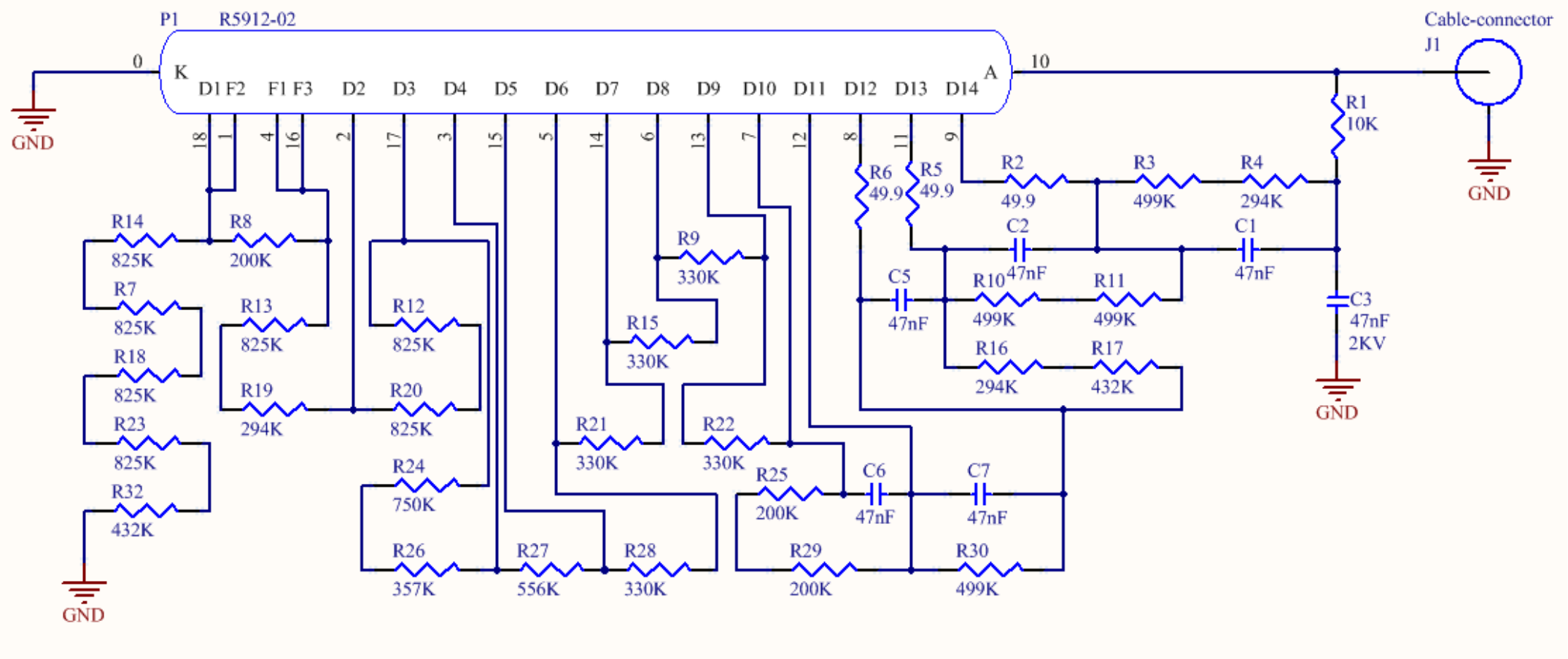}\label{fig:PB}}
\subfigure[]{\includegraphics[width=0.75\textwidth]{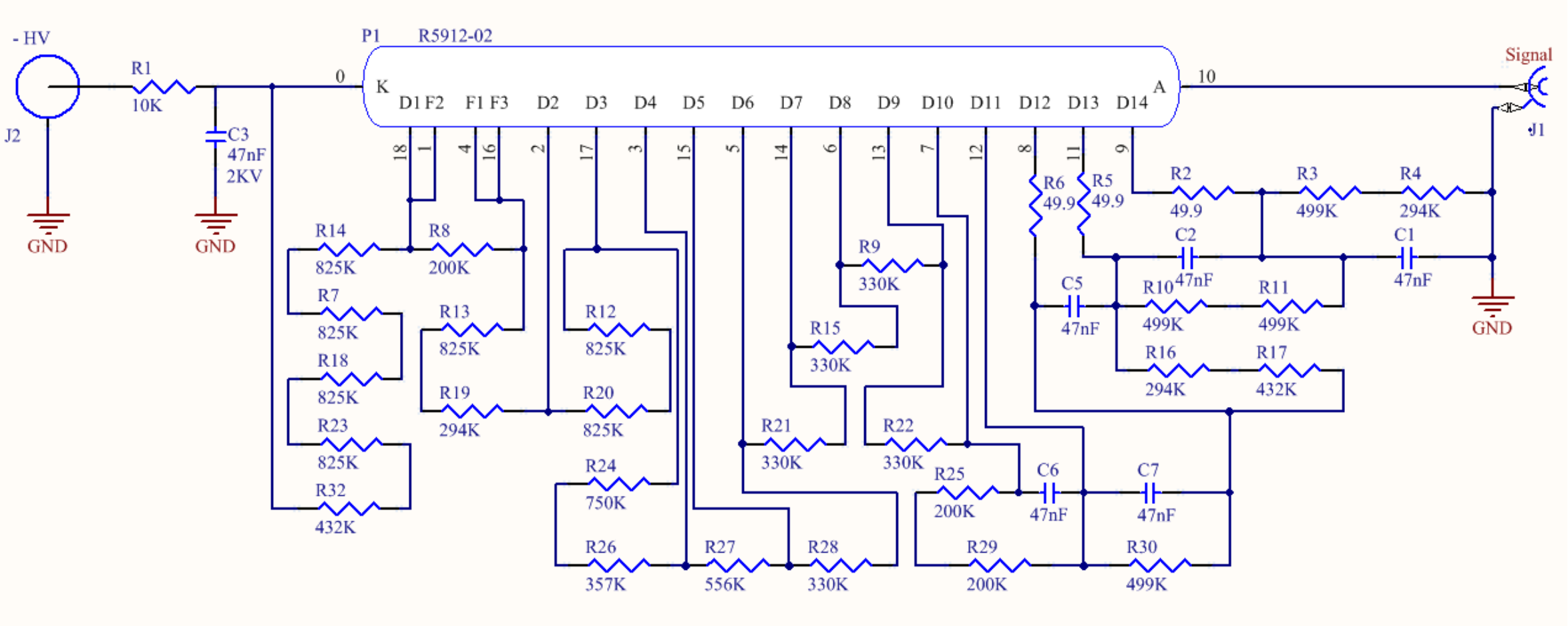}\label{fig:NB}}
\centering \caption{Diagrams of the two bases considered: (a) positive base and (b) negative base.}
\label{fig:basediagram}
\end {figure}

The PMT base was designed with passive components due to the degraded performance of semiconductors on cryogenic conditions. The components were carefully selected and tested in order to minimize their variations with the temperature. The circuit resistors were chosen following the voltage ratio recommended by the manufacturer to increase the linearity range. 
Resistors used on the base are SMD 0805 thin film from Vishay TNPW e3 series\footnote{www.vishay.com} and the capacitors are MLCC 1812 C0G dielectric from TDK C series\footnote{www.tdk.com}, except the high-voltage (HV) filtering capacitor (C3 in Figure~\ref{fig:basediagram}) that is a Polypropylene capacitor from Kemet R76 series. The total resistance of the circuit was set to 13.4\,M$\Omega$ with an average current $\sim$100\,$\mu$A (depending on the required voltage for each PMT), and the power dissipated per PMT base is $\sim$0.14\,W.

\begin {figure}[ht!]
\includegraphics[width=0.75\textwidth]{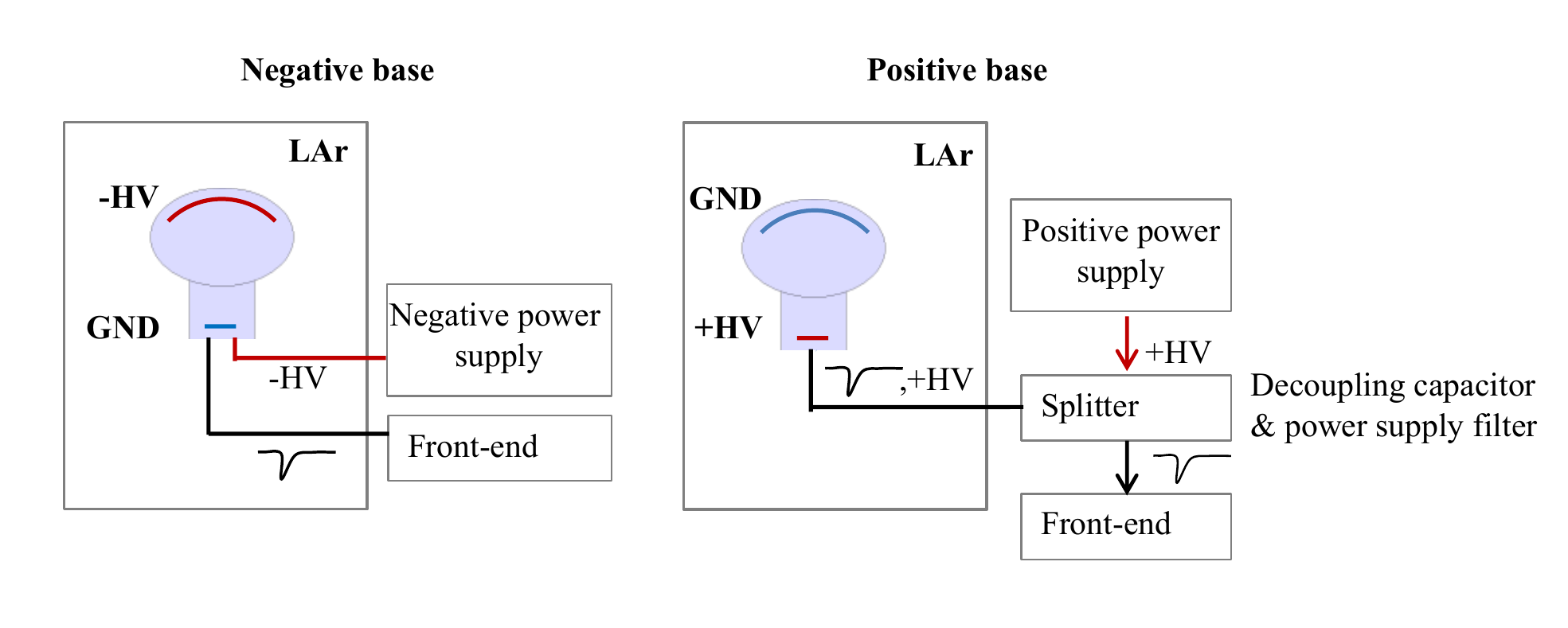}
\centering \caption{Schematic drawing of the two HV divider options for the base.}
\label{fig:bases}
\end {figure}

In order to choose the optimal PMT base design for ProtoDUNE-DP, two posible voltage configurations (positive and negative bias) were considered, see Figures~\ref{fig:bases}. In the so called positive base (PB)~\cite{microboone1}, see Figure~\ref{fig:PB}, a positive HV is applied at the anode and the photocatode is grounded, which reduces the noise; whereas in the negative base (NB)~\cite{clean1, icarus}, see Figures~\ref{fig:NB}, a negative HV is applied at the cathode and the anode is grounded. In the negative bias configuration, the photocathode is connected to HV, and special care must be taken to prevent spurious pulses due to HV leakage through the glass tube envelope to nearby grounded structures. The NB configuration requires two cables, one for HV and the other for the signal readout. Nevertheless, in this configuration it is more easily to read the PMT signal as it is referenced to ground. On the other hand, one advantage of the PB is that only one coaxial cable is required to carry the positive HV and to receive the signal from the PMT, but a decoupling circuit is needed to split the HV and the PMT signal. A dedicated splitter circuit was designed at CIEMAT to perform this function out of the cryostat. The splitter affects the real voltage value that is sent to the PB which is expected to be $\sim$7\% lower than the one read on the power supply. After some validation measurements, see section \ref{sec5.5}, the PB design was selected for ProtoDUNE-DP, as the total number of cables and feedthroughs in the detector is reduced and its behavior in terms of linearity and dark current is slightly better than the NB. All results presented in this study were achieved using the PB design with the exception of the results discussed in section~\ref{sec5.5}.

Once the PB was selected, all the bases were assembled, cleaned and tested at CIEMAT in air, Ar gas and liquid nitrogen (LN$_2$). Two tests were performed to the bases before being soldered to the PMTs and tested at CT: a resistance value of 13.4$\pm$0.1\,M$\Omega$ was confirmed, and 2000\,V were applied to the bases in Ar gas to verify the absence of sparks.


\section{Testing set-up}
\label{sec3}

\begin {figure}[ht]
\includegraphics[width=0.8\textwidth]{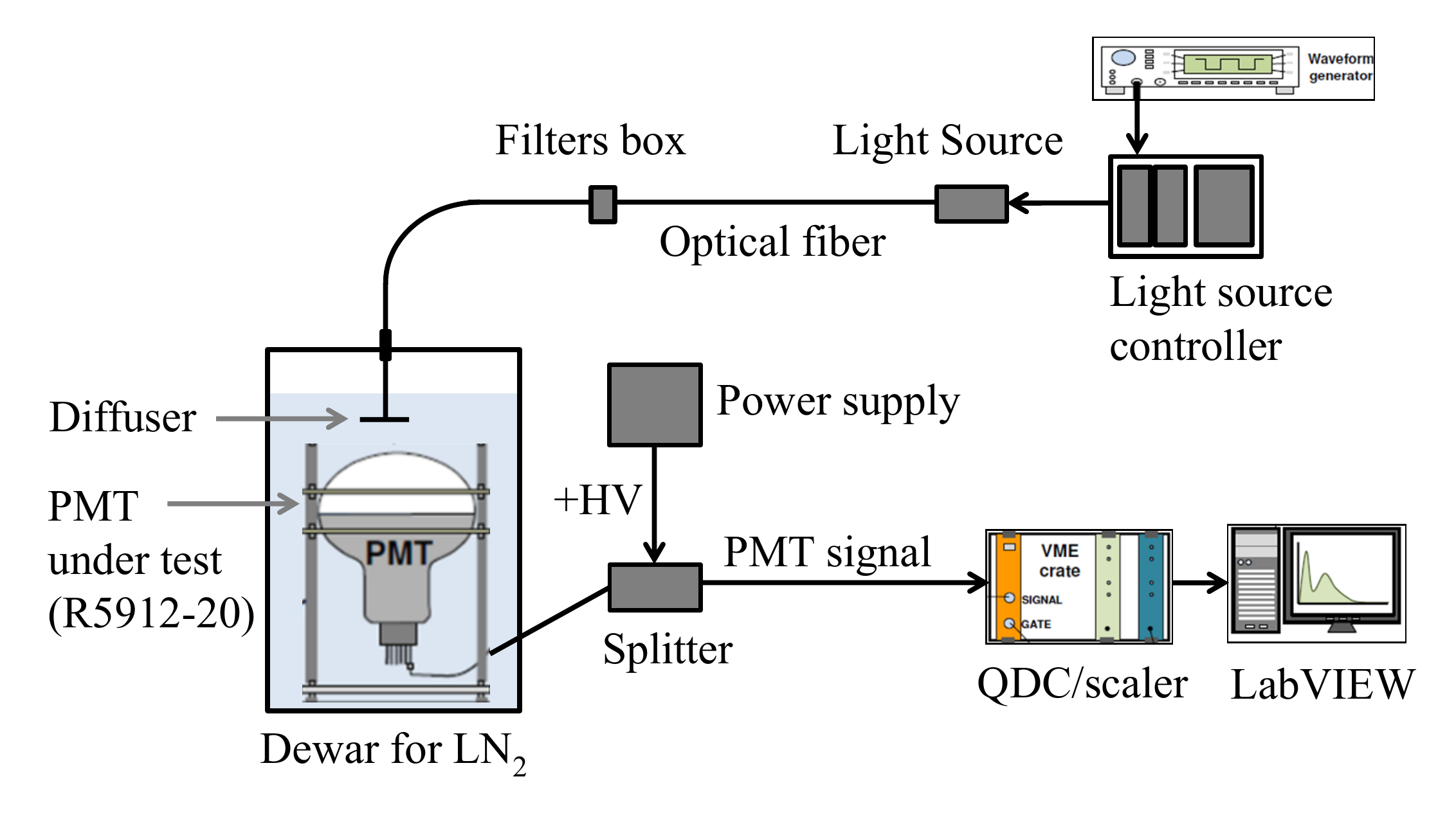}
\centering \caption{Schematic drawing of the PMT testing experimental set-up.}
\label{fig:setup}
\end {figure}

\begin {figure}[ht]
\subfigure[]{\includegraphics[height=0.25\textheight]{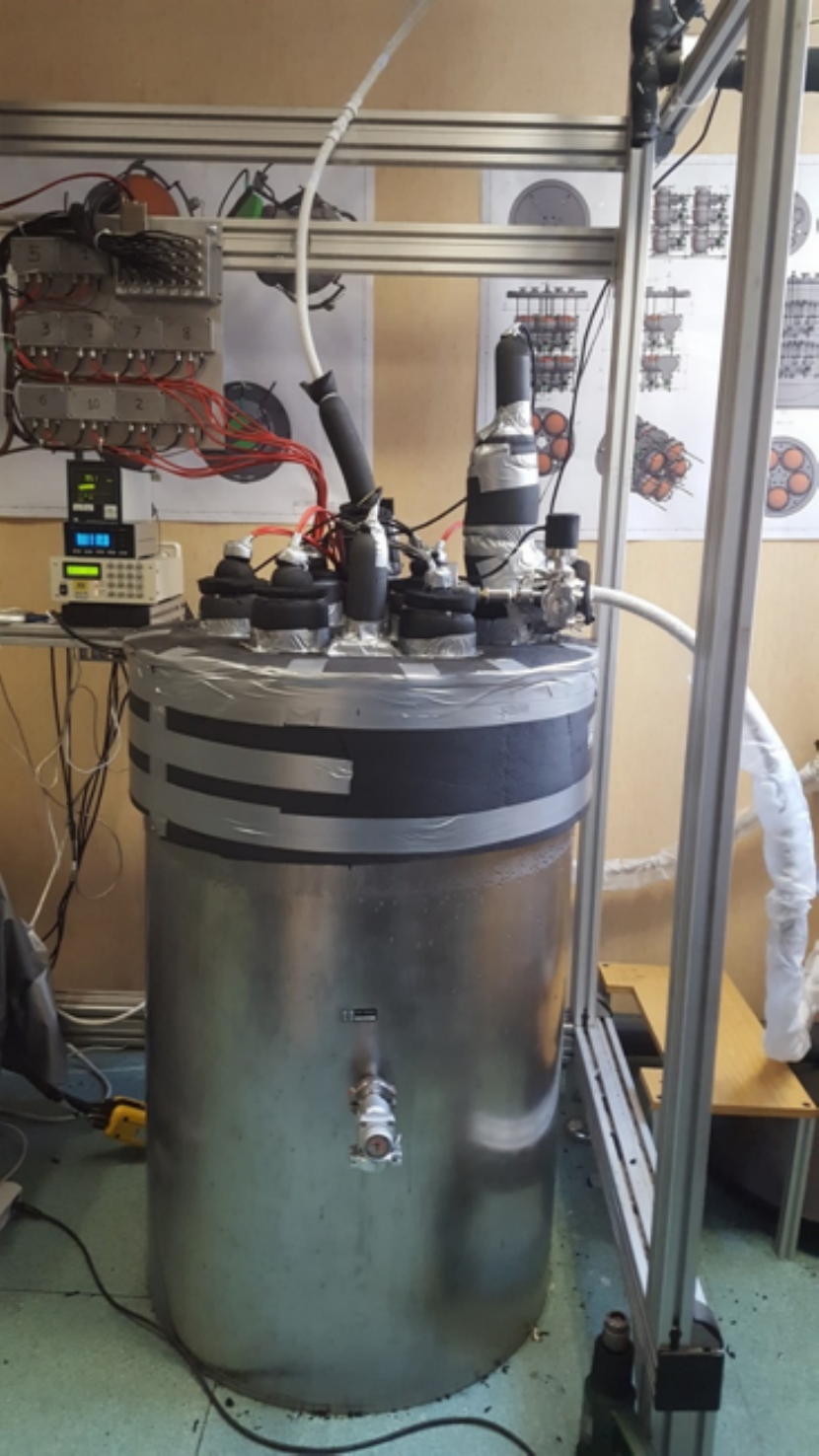}\label{fig:setup2a}}
\subfigure[]{\includegraphics[height=0.25\textheight]{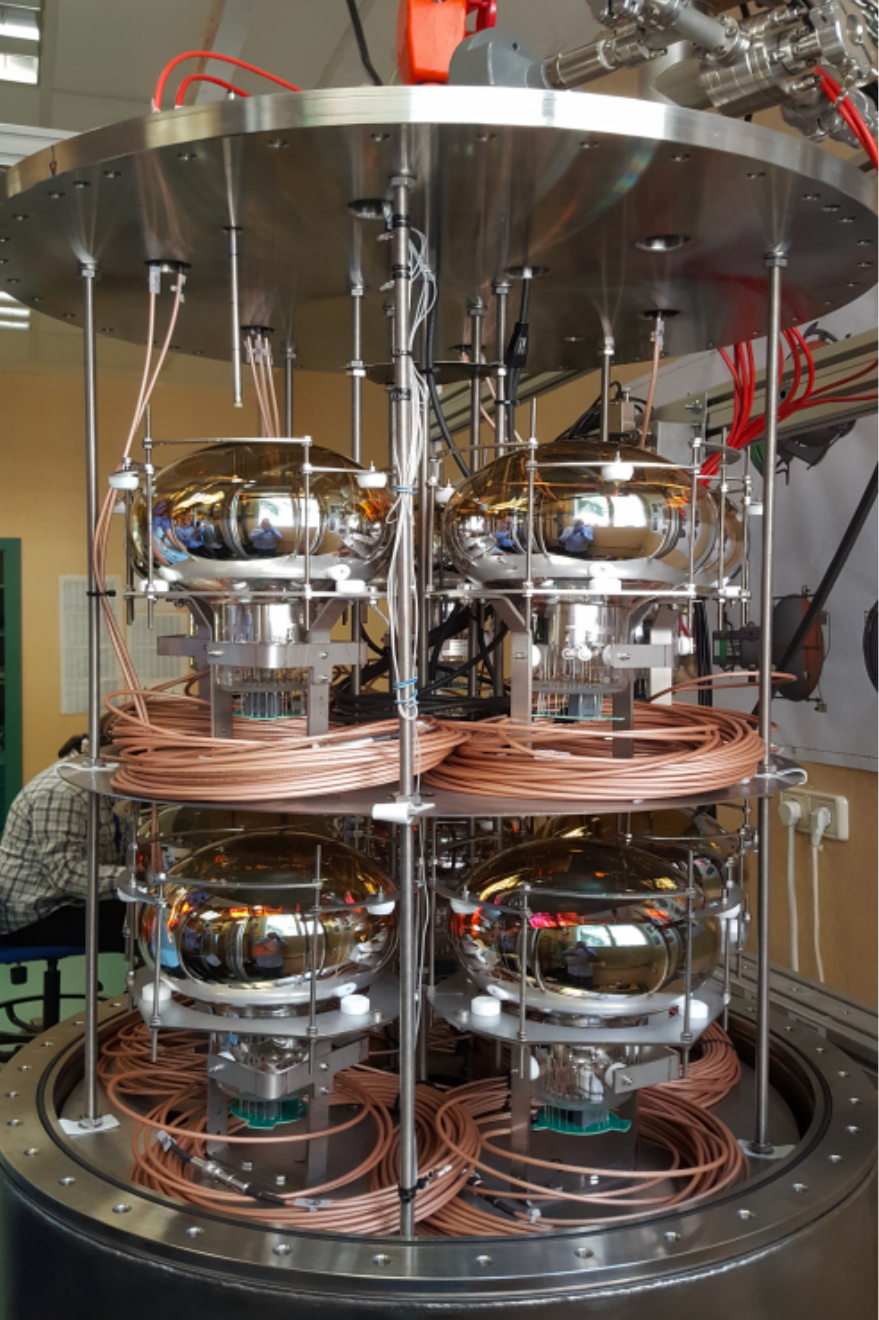}\label{fig:setup2b}}
\subfigure[]{\includegraphics[height=0.25\textheight]{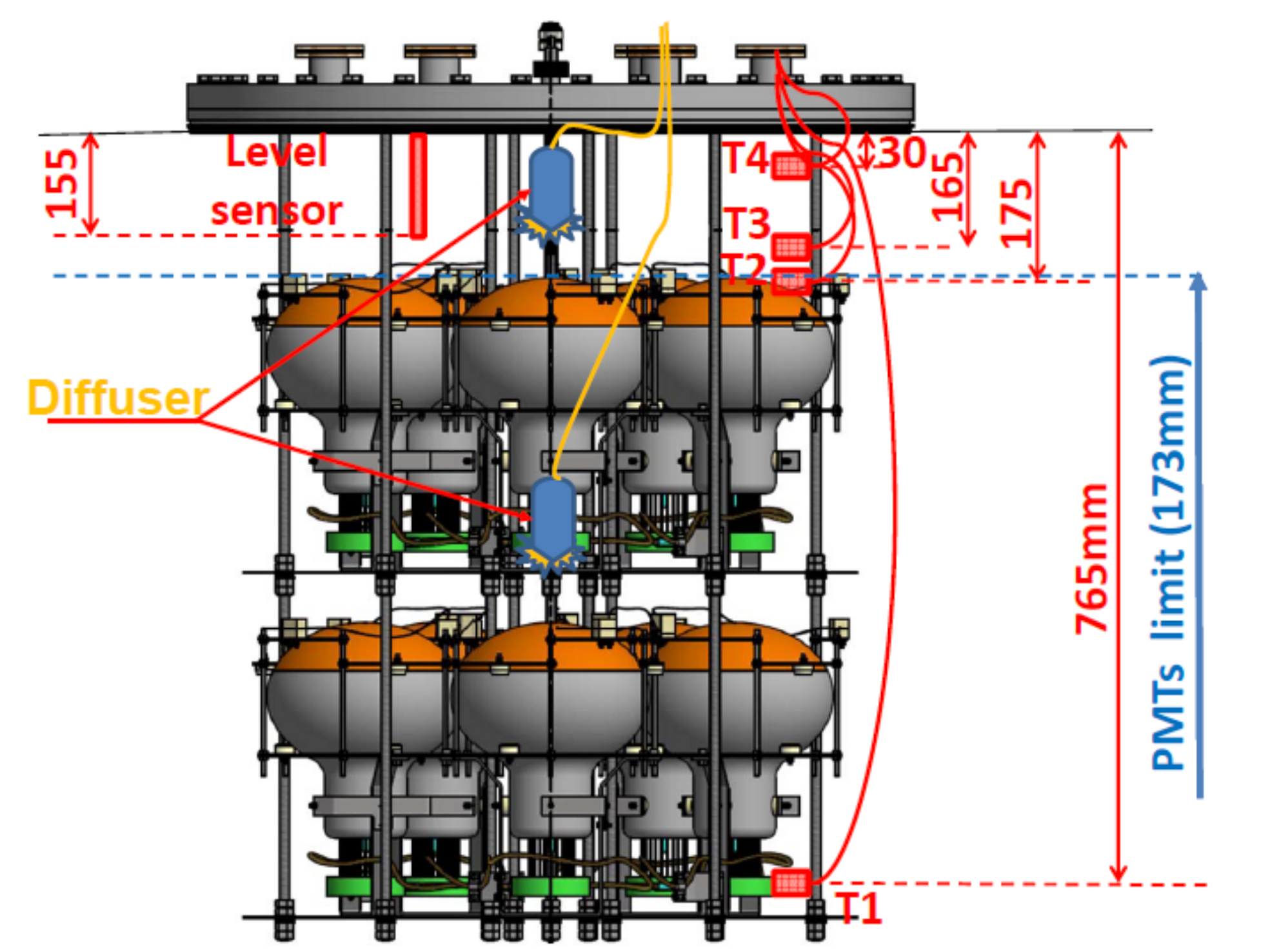}\label{fig:setup2c}}
\centering \caption{(a) 300\,L vessel used in the testing set-up. (b) 10 PMTs being installed in the vessel. (c) Schematic drawing showing the distribution of the 10 PMTs, temperature sensors (T1-T4), diffusers, and level sensor inside the vessel. }
\label{fig:setup2}
\end {figure}

A dedicated test bench was designed at CIEMAT for the PMT characterization. The measurements are performed to 10 PMTs at the same time inside a 300\,L vessel at RT and filled with LN$_2$ at 77\,K for the CT tests. A schematic drawing is shown in Figure~\ref{fig:setup}.  The vessel set-up and 10 PMTs installed are shown in Figure~\ref{fig:setup2}.

\begin {figure}[ht]
\includegraphics[width=0.7\textwidth]{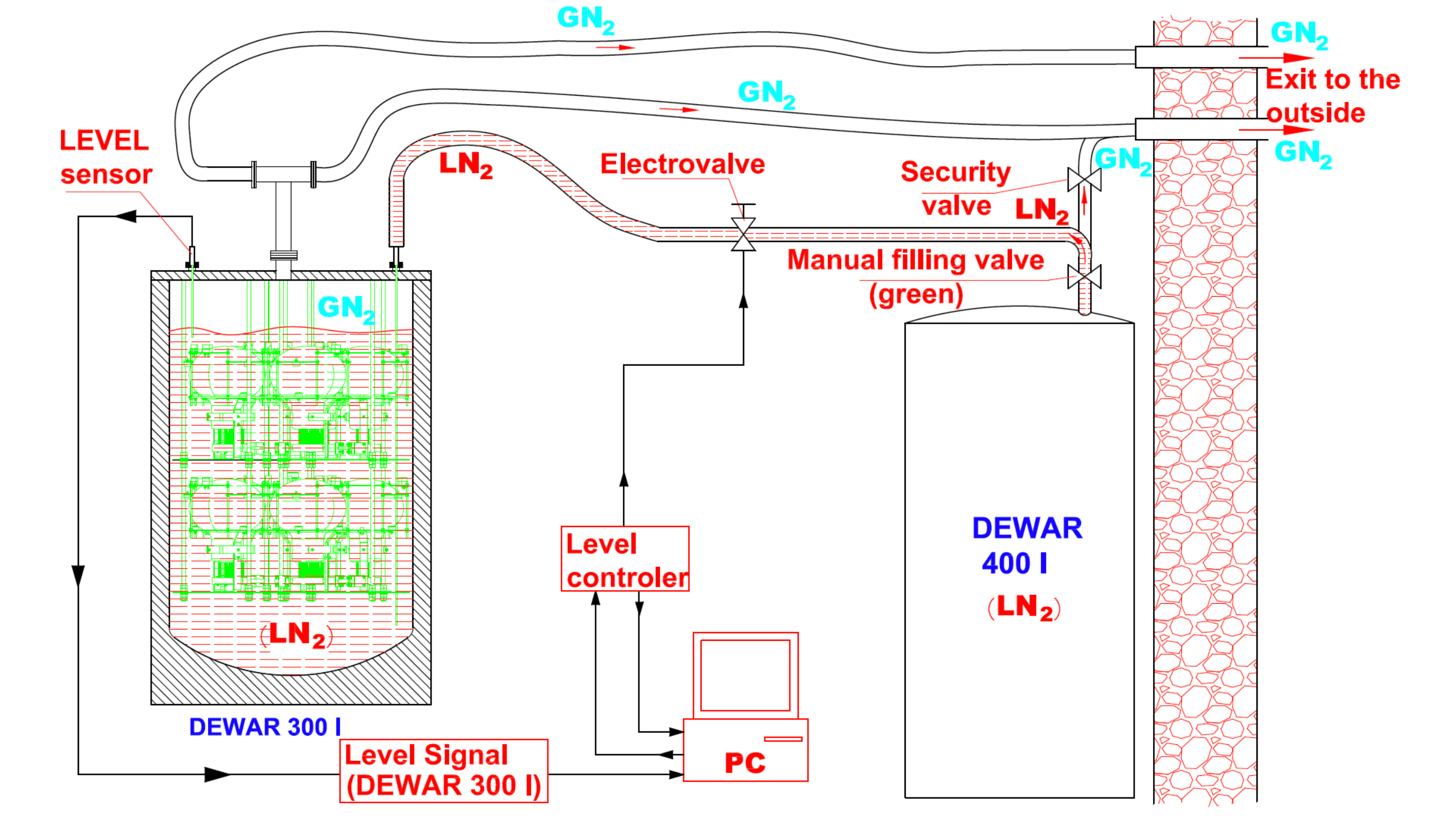}
\centering \caption{Detailed cryogenics set-up for 10 PMT testing.}
\label{fig:setup3}
\end {figure}

A 400\,L tank supplies LN$_2$ at $\sim$2\,atm pressure through a pipe directly to the 300\,L vessel where the PMTs are located at ambient pressure. A diagram of this system is shown in Figure~\ref{fig:setup3}. The system is automatically filled, using electro-valves controlled by level probes and temperature probes through a PC. The 10 PMTs are distributed in two levels, five per level, fixed to the lid through an internal structure, see Figure~\ref{fig:setup2c}. The cables of the PMTs, temperature and level sensors, and optical fibers pass through several CF40 ports. Each PMT is fixed to the internal structure with two M6 screws using the same support as it will be done in ProtoDUNE-DP. The PMTs are connected to the vessel feed-through by means of the same 21\,m cable that will be used on the final installation. The opening and closing of the lid is carried out with a little crane.

The main PMT features to measure are dark current, gain, and linearity. For each feature, a different electronics set-up is arranged using VME and NIM modules. To perform the dark current acquisition, a V895 discriminator and a V560E scaler tandem from  CAEN\footnote{http://www.caen.it/} is used. In order to calculate the gain and study the PMT response to light, the PMT output is measured with a V965A CAEN Charge-to-Digital Converter (QDC) in a 200\,ns window. The PMTs are biased using CAEN's N1470 power supply. The DAQ is remotely controlled with the aim of automating the data acquisition with LabVIEW software\footnote{http://www.ni.com/}. 

\begin {figure}[ht]
\subfigure[]{\includegraphics[width=0.45\textwidth]{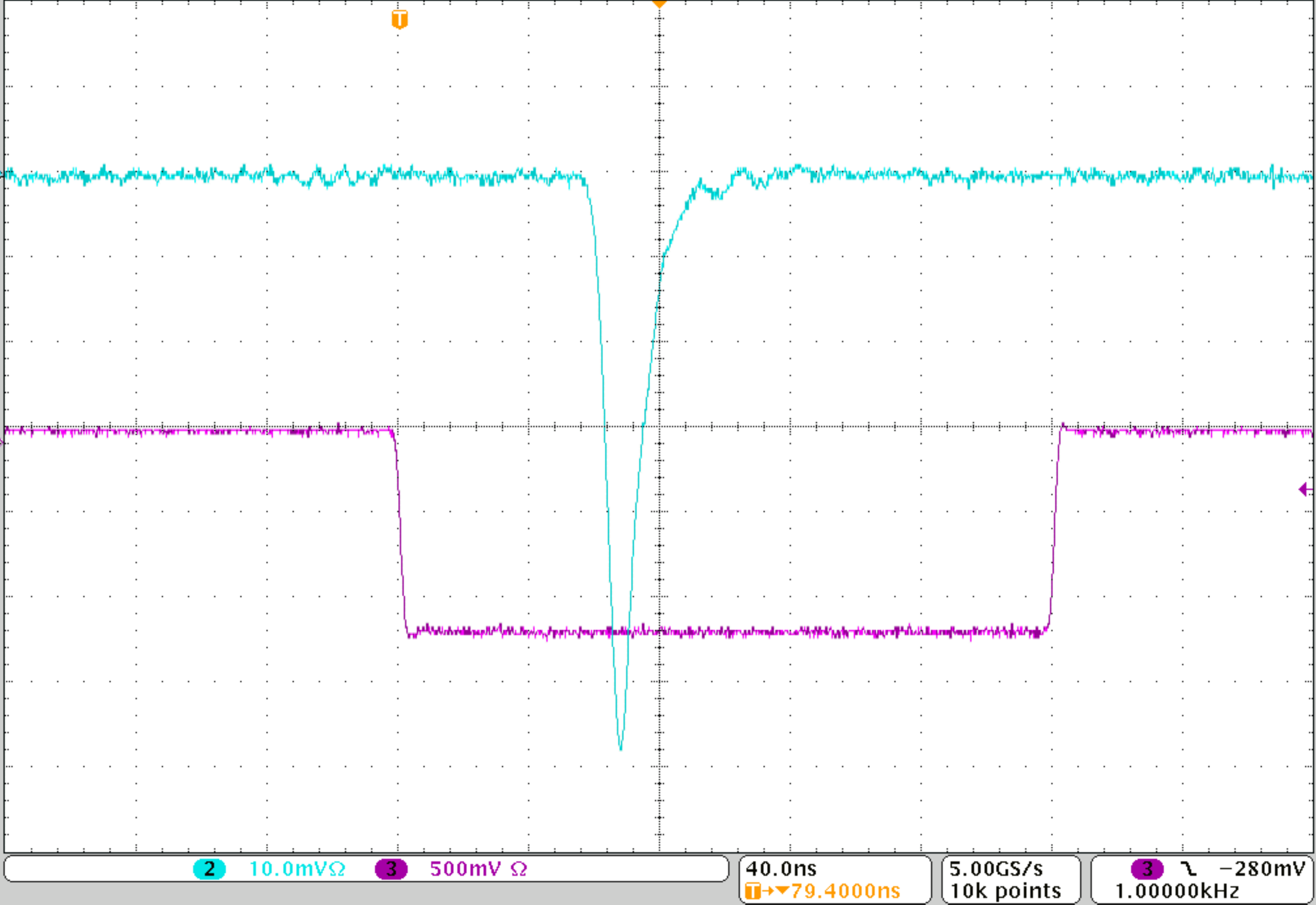}}
\subfigure[]{\includegraphics[width=0.45\textwidth]{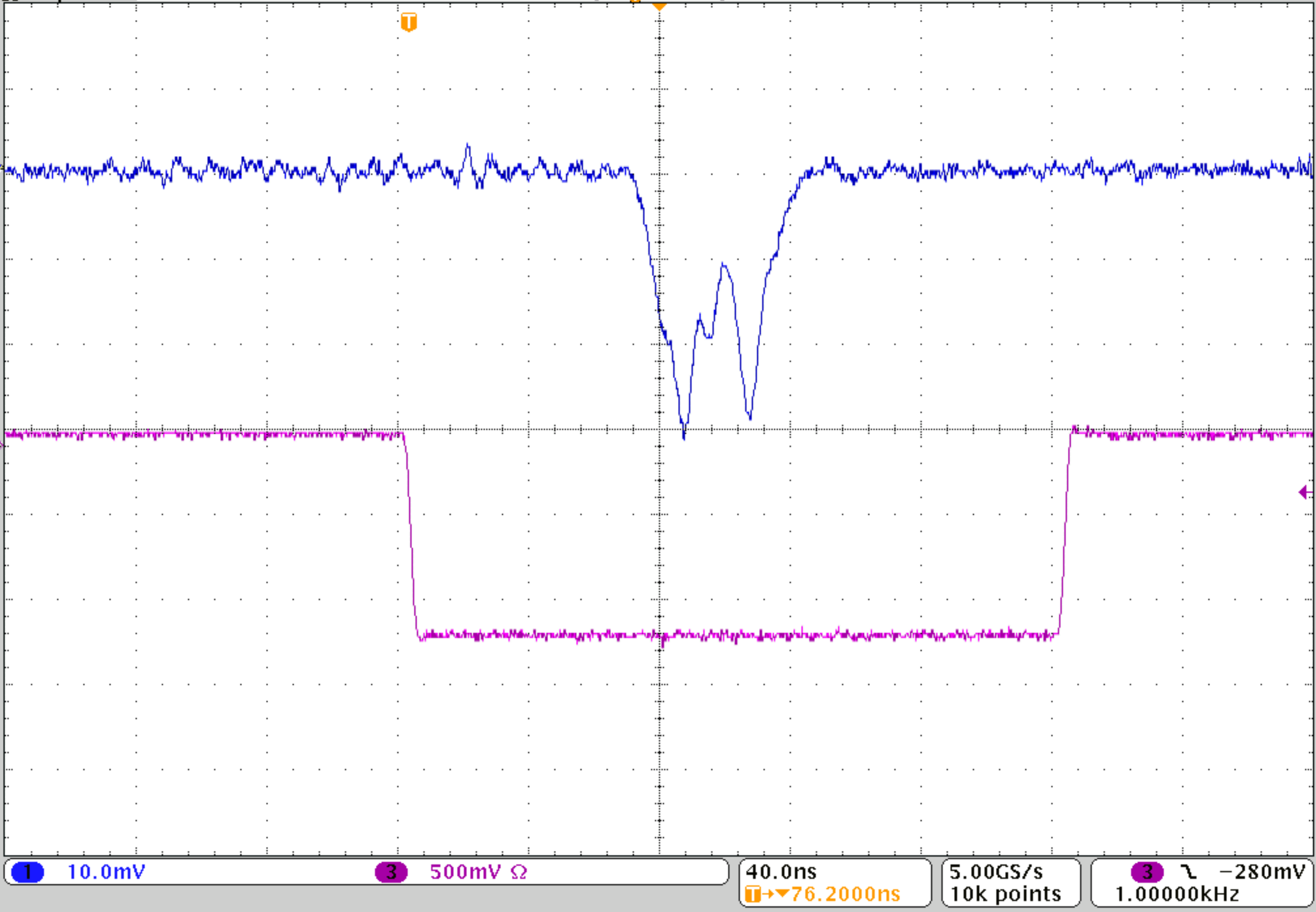}}
\centering \caption{PMT response to a $\sim$10\,p.e. light input provided by the laser (a) and LED (b). The integration window is shown in purple.}
\label{fig:sources}
\end {figure}

The light sources used are a PicoQuant GmbH\footnote{https://www.picoquant.com/} laser head, with a 405\,nm wavelength and a pulse width less than 500\,ps FWHM, and an LED pulser, with a 460\,nm wavelength and a pulse width of $\sim$40\,ns. An example of the PMT response using the laser and the LED is shown in Figure~\ref{fig:sources}. The QDC integration window and the light source are synchronized using a two output signal generator. The light is driven inside the dewar using hard clad silica multimode optical fibers from Thorlabs\footnote{https://www.thorlabs.com/}. The amount of light is tuned using a set of UV optical filters and the light is diffused in the detector volume. 

\section{PMT measurement methodology}
\label{sec4}

In this section, the approach and procedures used in the measurements, and the expected results are detailed. The studies carried out to fully understand the PMT behavior at CT include the dark current, the gain, and the linearity.

\subsection{Dark Current}
\label{sec4.1}

\begin {figure}[ht]
\includegraphics[width=0.49\textwidth]{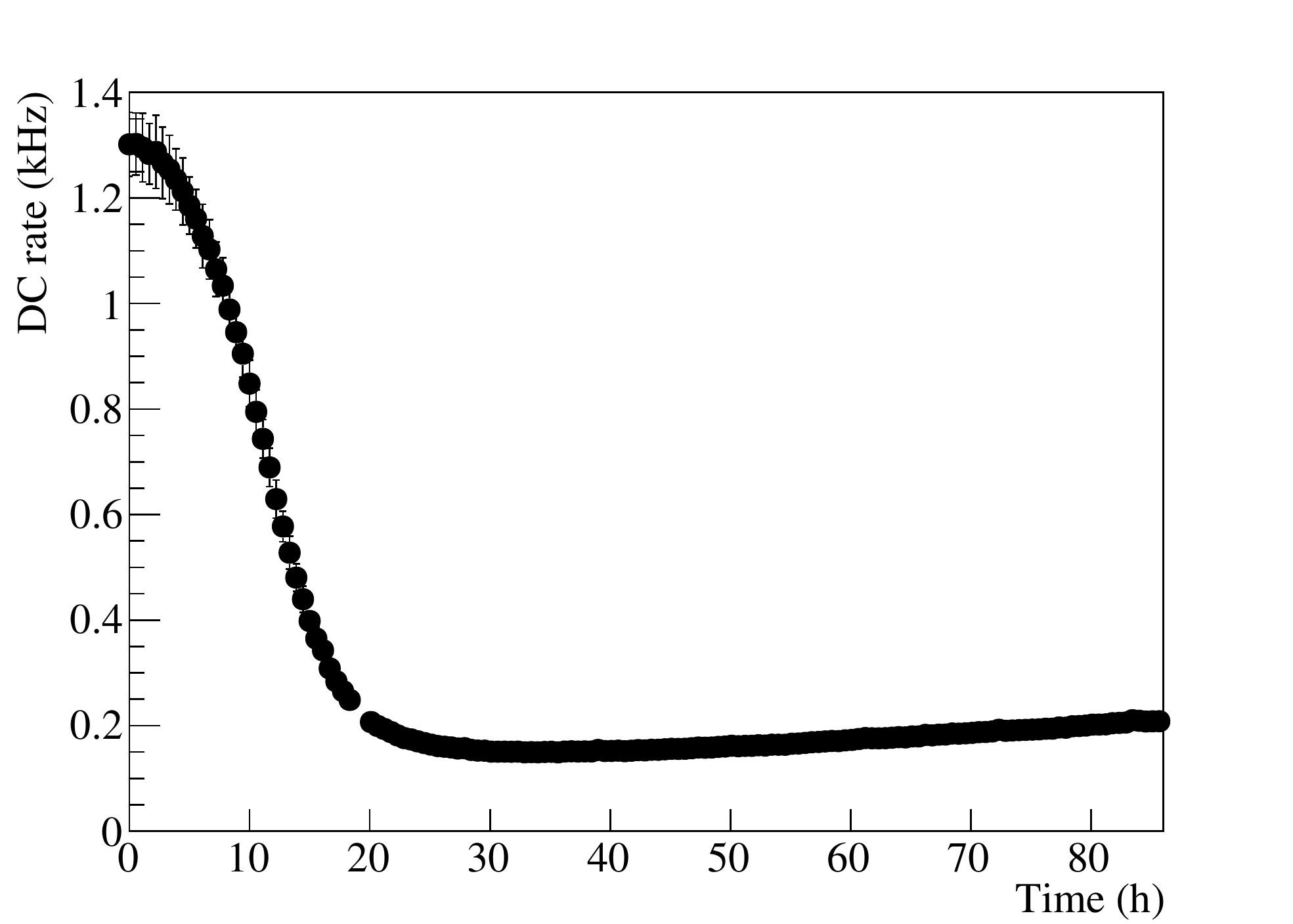}
\centering \caption{Evolution of the DC rate right after immersing a PMT in LN$_2$ and closing the vessel lid. DC measurements are taken every 10\,s and averaged every 13\,min in the plot (error bars are the rms).}
\label{fig:dcvstime}
\end {figure}

The dark current (DC) rate is the response of the PMT in absence of light. It is known that the main contribution to the DC at RT is the thermionic emission~\cite{ham2}. However, a non-thermal contribution increases the DC rate at CT \cite{meyer2,meyer3} . The DC rate is estimated as the average rate of detected signals larger than 7\,mV, which assures single photo-electron (SPE) triggering at an operating gain of 10$^7$ in a completely dark state. At RT, measurements are taken after at least 15 hours of complete darkness inside the vessel.  The DC evolution with time right after immersing the PMTs in LN$_2$ and closing the vessel is shown in Figure~\ref{fig:dcvstime}. It is clearly observed that initially the DC rate drops by more than one order of magnitude as the PMT was exposed to ambient light before the measurement. Then, the DC rate comes to a stable value with a small increase corresponding to the gain increase (see Figure~\ref{fig:gvstime}). As the PMT needs time to stabilize, DC measurements at CT are taken after $\sim$3~days. The same behavior is observed for all the tested PMTs.

The DC as a function of the HV is measured in 100\,V steps from 1000\,V to 1900\,V at RT and CT. Additionally, as the PMT manufacturer provided a reference value of the DC rate for every PMT at a gain of 10$^9$ at RT, a similar measurement of the DC at that gain is also performed. The conditions defined by Hamamatsu to measure the DC rate differ from the ones explained above, as the trigger is made at SPE level for a 10$^9$ gain, which is a higher threshold than the one used at CIEMAT.

\subsection{Gain and fatigue effect}
\label{sec4.2}

\begin {figure}[ht!]
\includegraphics[width=0.49\textwidth]{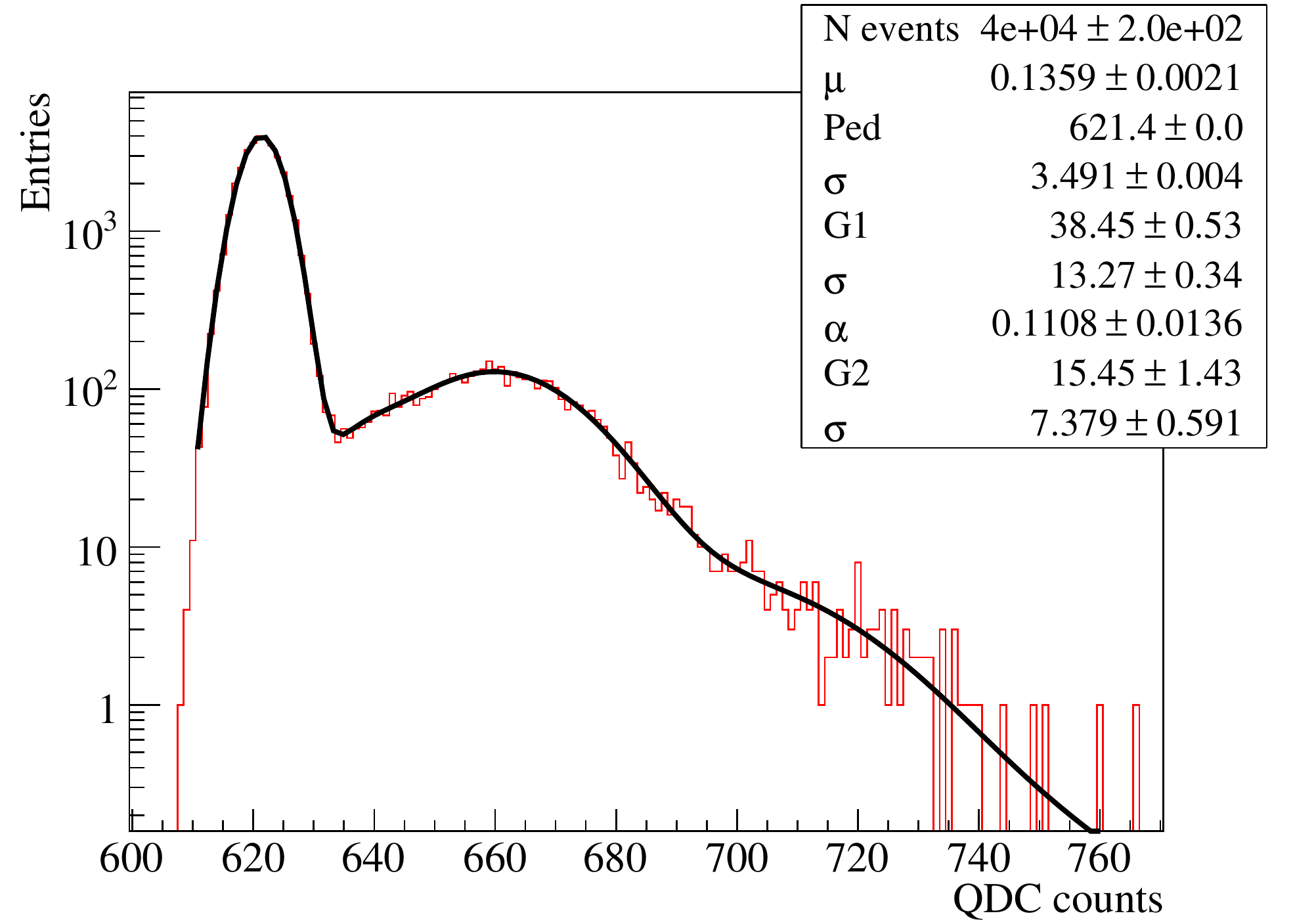}
\centering \caption{SPE spectrum (red) of a PMT at CT at 1500\,V, and fit results (black) where $\mu$ is the mean number of p.e. $Ped$ the position of the pedestal, $G1$ is the amplification of a p.e. from the first dynode (expressed as the mean pulse area), $G2$ from the second dynode, and $\alpha$ is the proportion of p.e. amplified by the second dynode.}
\label{fig:spe}
\end {figure}

To characterise the gain of each PMT, a small amount of photons is sent to the PMT to obtain the SPE spectrum. The SPE spectrum is fitted to a convolution of a Poisson distribution, which models the number of p.e. generated in the photocathode, and a binomial distribution considering two posible amplification paths: through the first dynode and directly starting in the second dynode~\cite{bellamy}. An example of the fit is shown in Figure~\ref{fig:spe}.

The gain vs HV, known as gain-voltage curve, is measured from 1100\,V to 1900\,V in 100\,V steps. Then, a fit is done following the power law $G= A V^B$ being $A$ and $B$ constants dependent on the number, structure, and material of the dynodes~\cite{ham2}. 

\begin {figure}[ht!]
\includegraphics[width=0.49\textwidth]{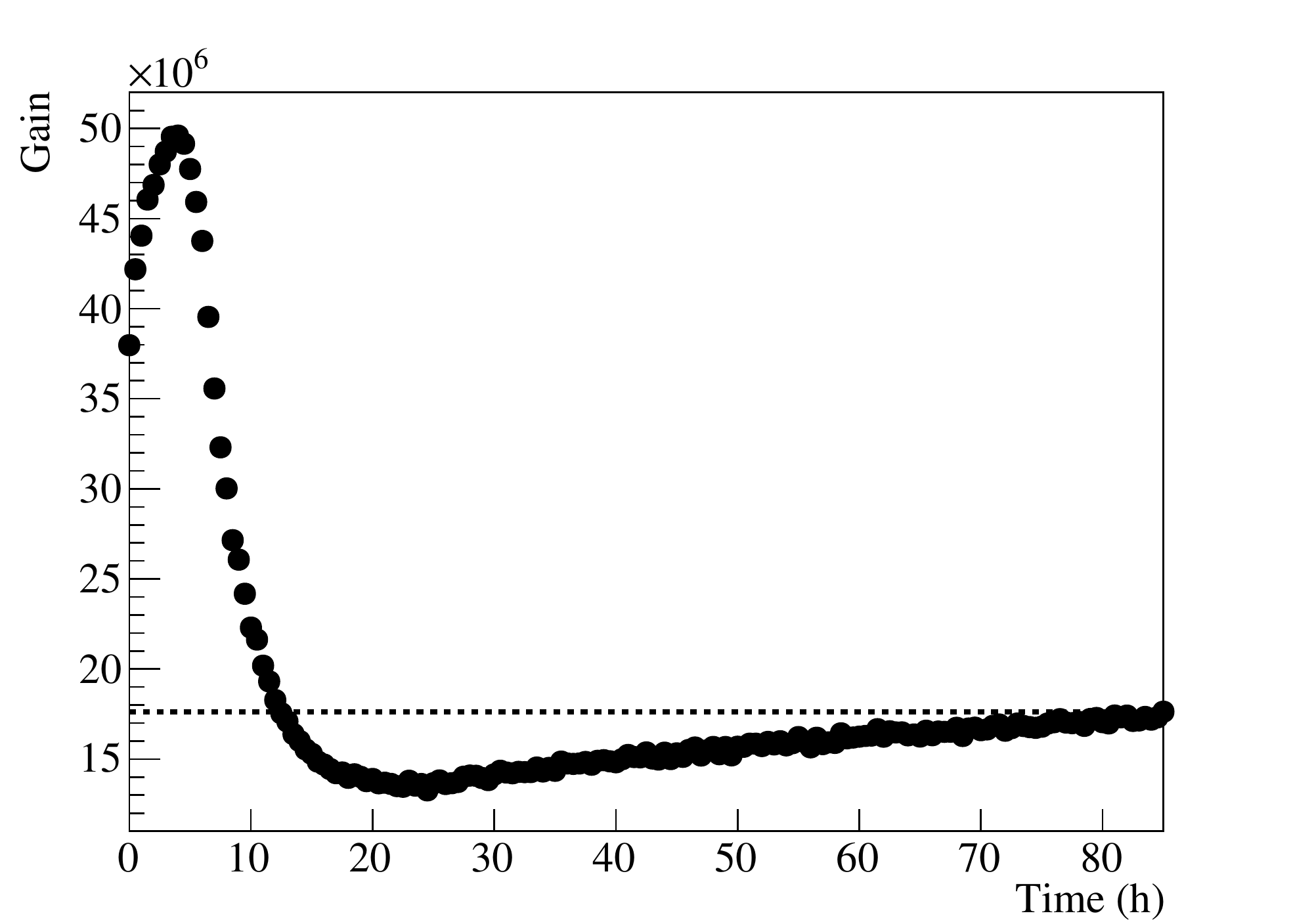}
\centering \caption{Evolution of the gain right after immersing a PMT in LN$_2$ and closing the vessel lid. Gain is calculated every 30 minutes. The dotted line represents the nominal gain obtained 3 days after immersing the PMTs in LN$_2$.}
\label{fig:gvstime}
\end {figure}

The gain evolution with time right after immersing the PMTs in LN$_2$ is shown in Figure~\ref{fig:gvstime}. During the first $\sim$5 hours the gain increases as the PMT vacuum improves, then it drops during $\sim$10 hours by a factor of 4 while the PMT is cooling down. Finally, a small drift takes place and gain reaches a stable value after $\sim$3~days. As a result, tests are taken always at least 3 days after immersion. A very similar behavior is reported in~\cite{microboone1}.

A variation of the gain is expected when the PMT output current increases, either, due to high gain, to high light intensity or to high rate of the light (or a combination of them). This variation of the gain is called fatigue~\cite{fatigue} and can distort the results of the measurements if the PMT cannot recover from this effect between measurements. At RT, no variation of the PMT gain is observed during the tests. The PMT fatigue recovery is much slower at CT allowing to observe a gain reduction. This gain reduction depends on the PMT output current generated during the previous test. In addition, once the cause of the gain reduction is over, the PMT requires a long time to recover the initial gain, and this time is also dependent on the amount of current generated in the PMT. Both factors, gain reduction and recovery time, vary from one PMT to another.

\begin {figure}[ht!]
\subfigure[]{\includegraphics[width=0.49\textwidth]{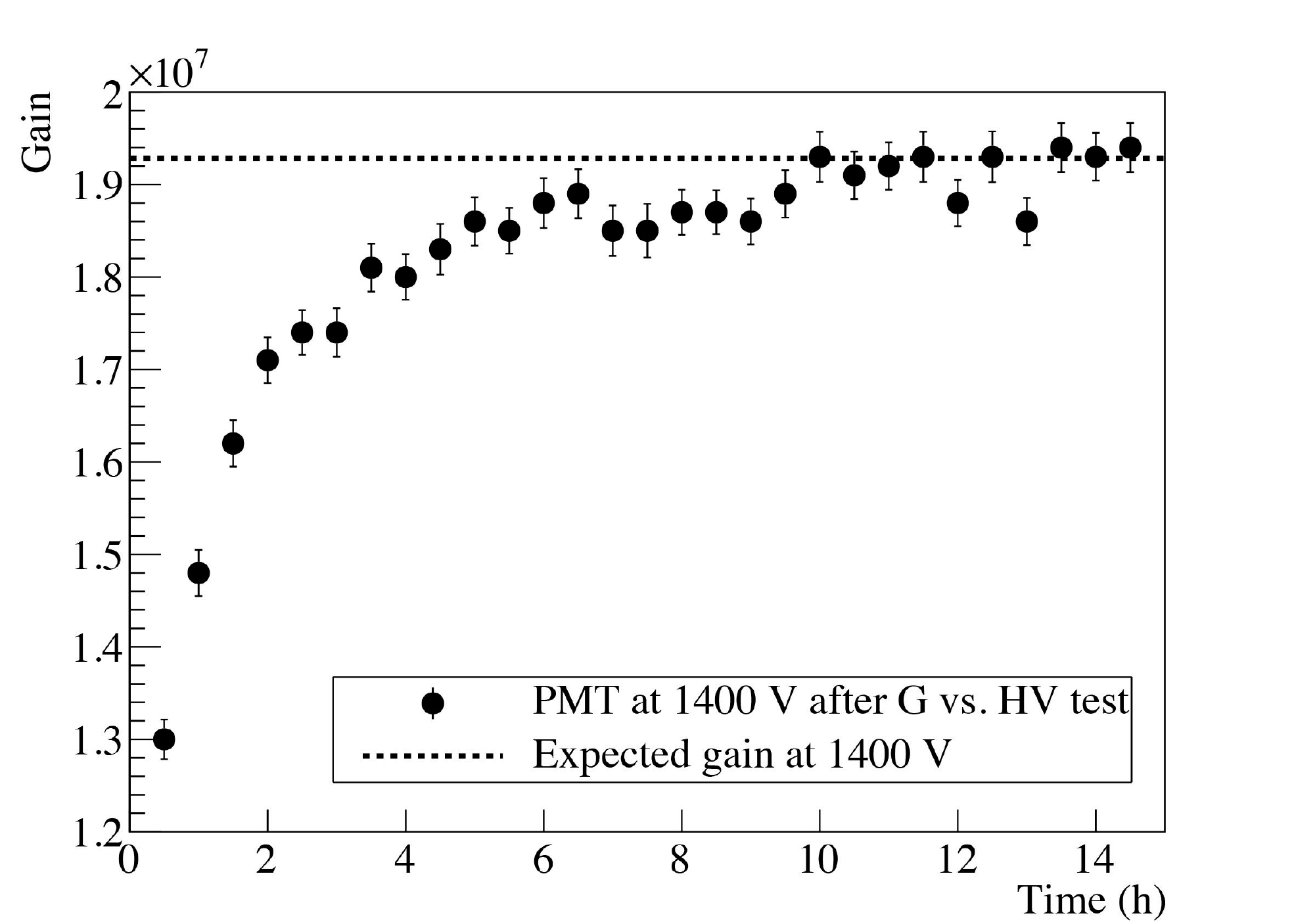}\label{fig:fatiguea}}
\subfigure[]{\includegraphics[width=0.49\textwidth]{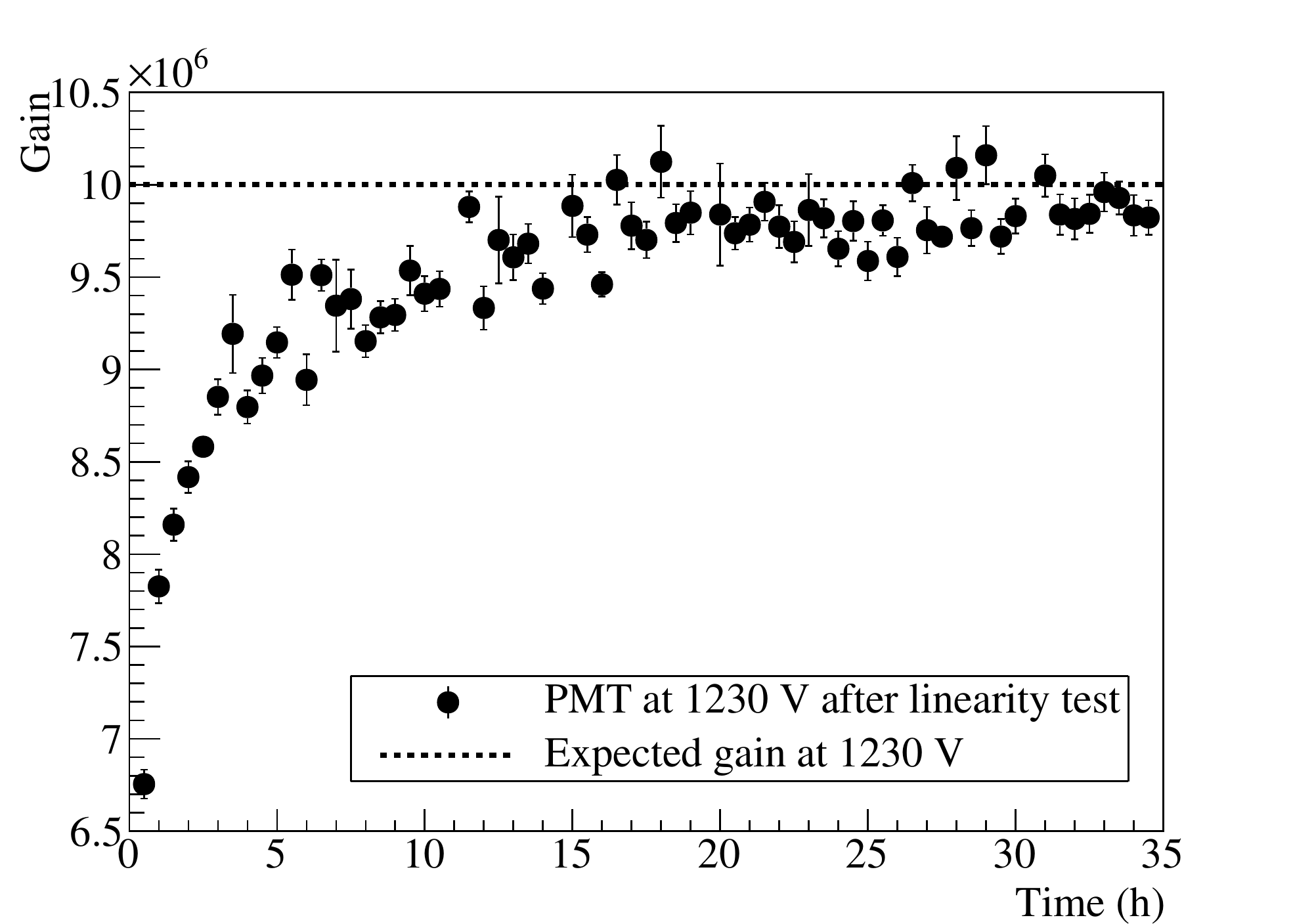}\label{fig:fatigueb}}
\subfigure[]{\includegraphics[width=0.49\textwidth]{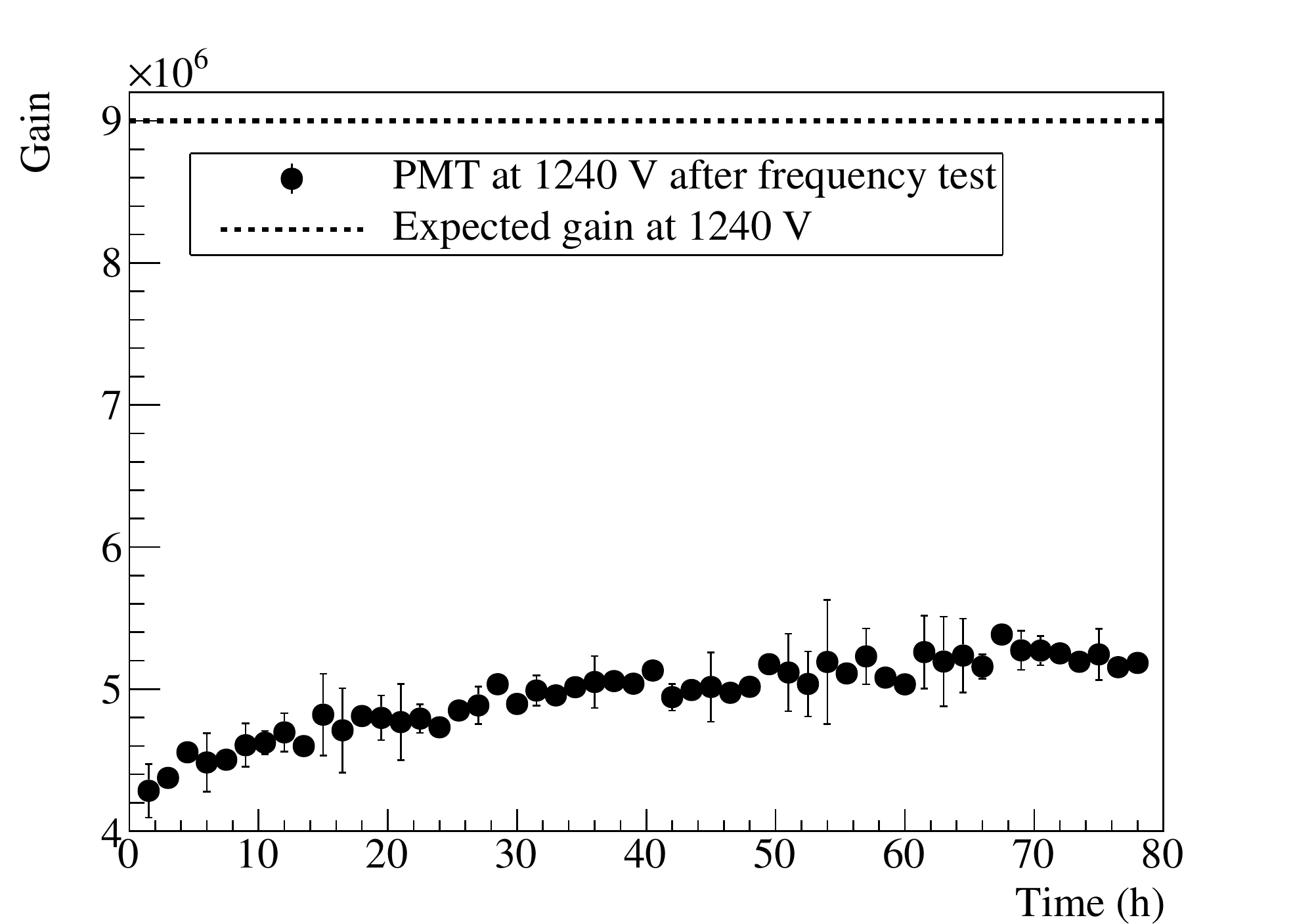}\label{fig:fatiguec}}
\centering \caption{ (a) Gain evolution at 1400\,V after raising the voltage to 1900\,V. (b) Gain recovery time after linearity measurements. (c) Gain recovery time after high rate measurements. The dotted lines represent the expected gain without the fatigue effect.}
\label{fig:fatigue}
\end {figure}

The PMT gain is monitored over time after the tests that produce a high PMT output current. The gain monitoring is performed triggering the PMTs with low light level (SPE) at a rate of 100\,Hz. First, the recovery time is studied at 1400\,V after gain vs HV measurements,  see Fig.~\ref{fig:fatiguea}. Although the average number of p.e. is lower than 1, at the maximum voltage (1900\,V) the output current is high enough to reduce the effective PMT gain by 30\% with respect to the initial measured gain at 1400\,V and the gain recovery requires about a day at CT. Second, after the linearity measurements where PMTs are exposed to light levels of $\sim$1000\,p.e. and up to a 10$^8$ gain, a similar behavior is observed, see Fig.~\ref{fig:fatigueb}. Finally, a more permanent effect over the gain is caused by high rate signals of the order of MHz, as can be observed in Fig.~\ref{fig:fatiguec}. In this case, the gain decreases by a factor of 2. For some PMTs, a week after illuminating them with high frequency signals, the gain is still 40\% lower than the nominal gain. However, other PMTs recover faster (in 3 days). In any case, the recovery time varies a lot from one PMT to another. The theoretical value of the PMT maximum output current during the previous test is calculated to verify that the gain recovery time depends on it. The PMT maximum output current is calculated as $I_{out} = e^- \cdot G \cdot npe \cdot f$, where $e^-$ is the electron charge, $G$ is the PMT gain, $npe$ is the average number of p.e. during the test, and $f$ the light pulse rate in Hz. The results are a few $\mu$A for the gain vs HV and linearity tests, and hundreds of $\mu$A for the light rate test, so  the maximum output current looks proportional to the observed recovery time.

\subsection{Linearity}
\label{sec4.3}

The PMT response is studied as a function of the light intensity. The amount of scintillation light arriving to the PMTs in ProtoDUNE-DP will vary from a few p.e. to thousands of p.e. depending on the particle track energy and distance to the PMTs. Thus, the response of the PMTs should be linear in a wide dynamic range. The PMT base was designed to fulfill this goal, and the deviation of the linear response is primarily caused by the anode saturation.

The PMT linearity is studied at gains up to $10^8$ using a set of 11 filters with different attenuation factors connected to the light source through optical fibers. To avoid the fatigue effect, explained in section~\ref{sec4.2}, measurements are done from low light (SPE) to high light levels (up to 1000 p.e.). The expected amount of light is estimated relatively to the measured one in the linear region,   taking three reference filters and considering their transmission factor. As known from previous tests, the PMT response is linear up to at least 100 p.e. for gain $\sim$10$^7$ and up to 40 p.e. for gain $\sim$10$^8$. For low intensity illumination (SPE regime) the average number of p.e. is obtained from the fit of the SPE spectrum, but it is not used as reference due to its large uncertainty. The 405\,nm laser (<1ns pulse) and the 460\,nm LED (40\,ns pulse) are used to check the effect of the pulse profile on the anode saturation. For the same amount of detected light, the shorter the light pulse, the higher output peak current leading more easily the PMT to saturation due to space-charge effects on the last dynodes.

\subsection{Light rate}
\label{sec4.4}

In ProtoDUNE-DP, a continuous background of light pulses is expected due to the secondary scintillation light produced on the gaseous phase of the argon by the drifted electrons. To study how this can affect the PMT performance, dedicated tests are carried out. In these tests, the light intensity is set to different levels, from few p.e. to 150 p.e., using the set of filters. For each light level, the light pulses emission frequency is swept from low (100\,Hz) to high frequencies (10\,MHz). The tests are carried out with the laser and the LED to check the effect of the pulse profile. For each combination of light level and pulsed frequency, the charge spectra of the PMT is obtained to calculate the averaged amount of light observed by the PMT.

Increasing the rate of the light pulses produces a proportional increase of the average output current. As the base design is based on resistors, the PMT inter-dynode voltages have a dependency with the PMT output current making the PMT response not linear when the light rate increases over a certain limit. The current through the base resistors has two components: the polarization current, constant and provided by the power supply, and the PMT output current that goes through the resistors in the opposite direction. As the total voltage applied to the base is kept constant by the power supply, a decrease on the last dynodes voltage (by the increase of the output current), makes higher the voltage on the first dynodes increasing the PMT gain. If the voltage on last dynodes continues decreasing (by the increase of the output current) at some point this voltage is not enough to maintain the current flow to the anode, and it decreases following the typical I-V curve of a vacuum diode. The only difference is that in the vacuum diode, the cloud of electrons is generated by a filament while in the PMT it is generated by the incident light and the previous dynodes. Besides that, the PMT output current does not depend any more on the light input intensity, only on the light rate that is changing the voltage on the last dynodes. 

Then, the expected PMT response vs. light rate can be divided in three zones: first, the linear response, from $<$1\,Hz; second, the over-linearity region, where the PMT gain increases; and third, the saturation region, when the voltage between the last dynode and the anode is close to zero and the PMT output decreases with the light rate. The PMT output is never zero because the initial velocity of the p.e. is not zero, therefore some of them can be collected by the last dynode even if the potential difference between the cathode and the first dynode is zero. To reach the zero output this potential should be negative.

At CT, the fatigue effect observed on the PMT gain (as explained in section~\ref{sec4.2}), compensates the beginning of the gain increase with the light rate leading to a small reduction of the over-linearity region.


\section{PMT validation results}
\label{sec5}

Although these PMTs are designed to operate at CT, the manufacturer does not provide information about the CT behavior. Then, to validate the PMT model selection, different tests were carried out to several PMTs at RT and CT, and the results are presented in this section. 

\subsection{Dark Current}
\label{sec5.1}

\begin{figure}[ht!]
\centering
\subfigure[]{\includegraphics[width=0.49\textwidth]{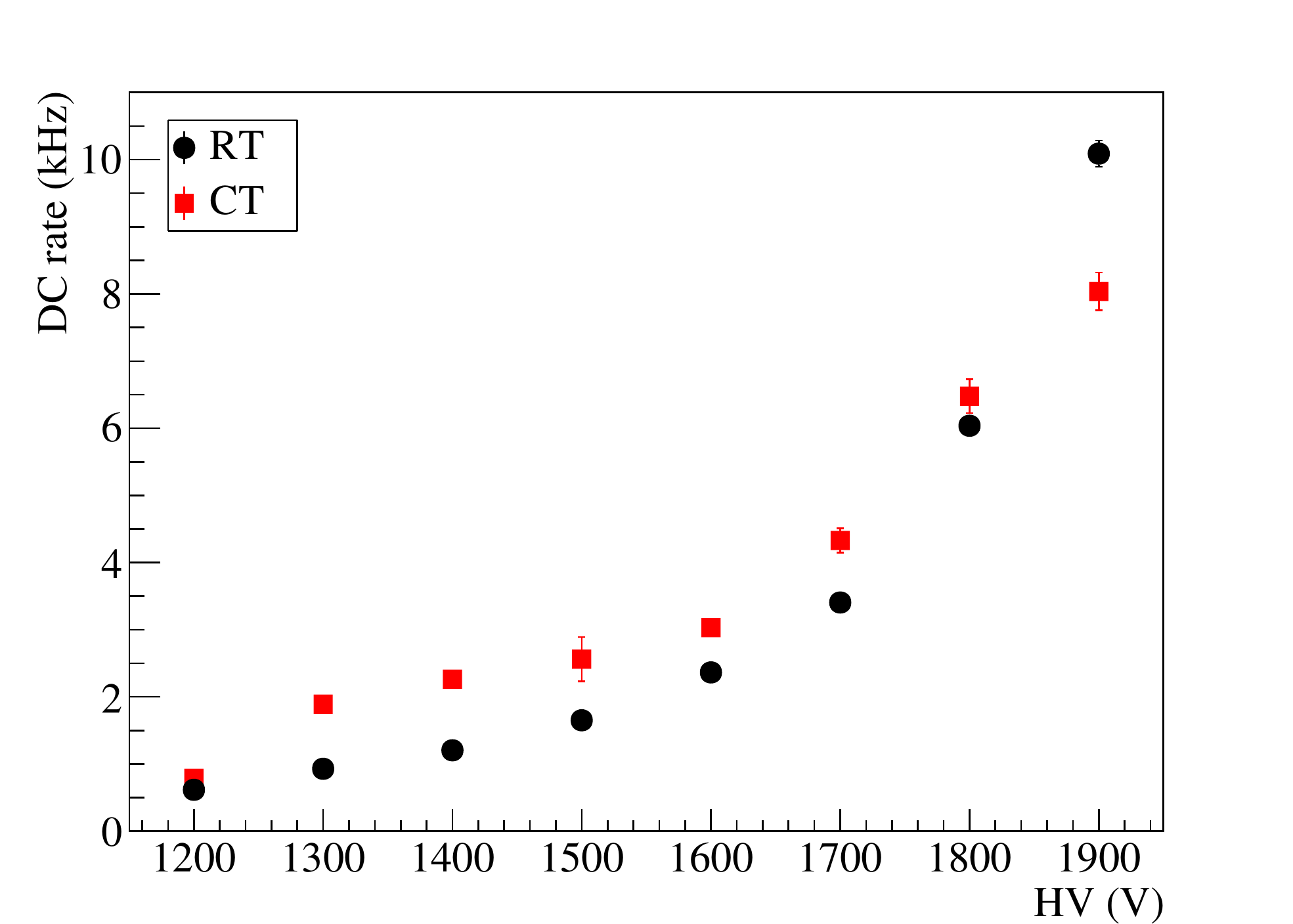}}
\subfigure[]{\includegraphics[width=0.49\textwidth]{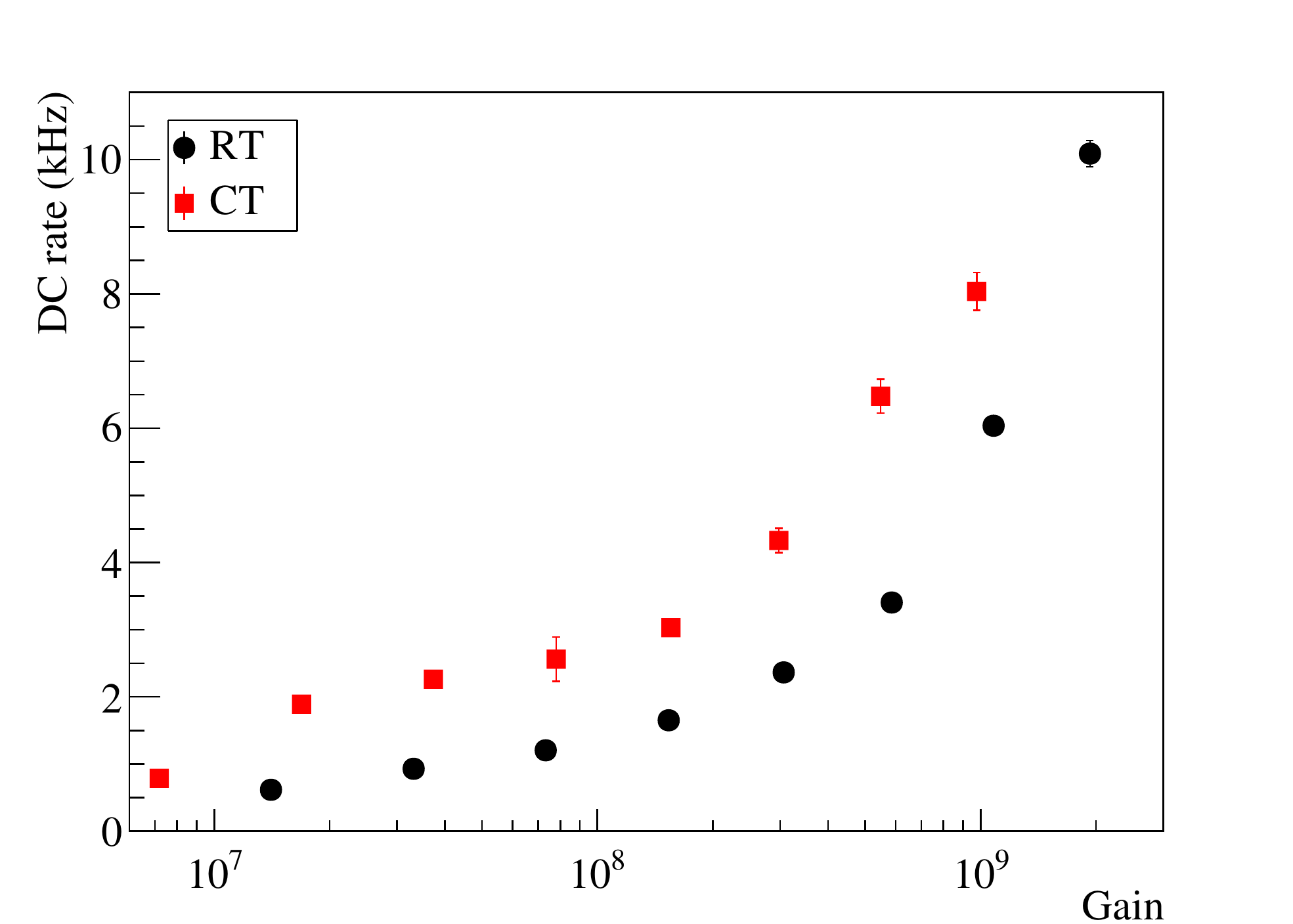}}
\caption{DC vs HV (a) and vs gain (b) at RT and CT for one PMT. For each HV, DC is measured every 10\,s during 5\.min, the average value is plotted and the error bars represent the rms.}
\label{fig:dc_vs_hv_rt_ct}
\end{figure}

The DC rate as a function of the HV is measured. Figure~\ref{fig:dc_vs_hv_rt_ct} shows a typical plot of the DC at RT and CT. The plot is also done as a function of the gain to equalize the response at RT and CT, as expected. In general, the DC at CT is higher than at RT for the same gain. For this particular PMT, the DC increases from 0.6\,kHz at RT to 1.9\,kHz at CT for a gain of 1.5$\cdot10^7$, being the behavior of this PMT representative of other PMTs. 

\subsection{Gain}
\label{sec5.2}

\begin{figure}[ht!]
\centering
\includegraphics[width=0.49\textwidth]{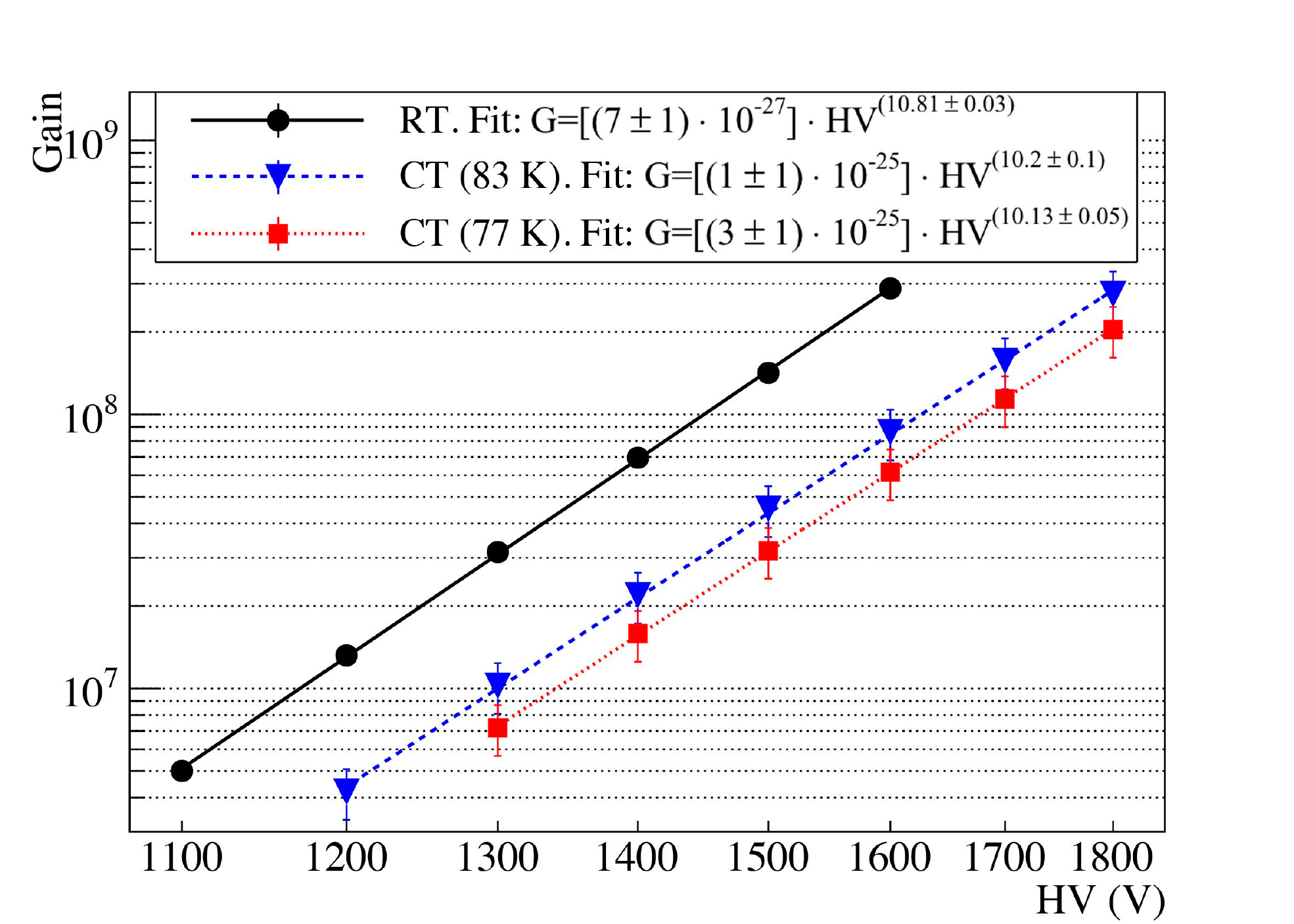}
\caption{Gain vs HV at RT, at two CT (77\,K and 83\,K) for the same PMT. The dots represent the measurements, and the lines the fit explained in section \ref{sec4.2}.  The vertical error bars correspond to the average variation in the gain obtained from repeated measurements: 7\% at RT and 21\% at CT.}
\label{fig:gain_vs_hv_rt_ct}
\end{figure}

Gain-voltage curves at RT and CT (at 77\,K) are measured, and the result for one PMT can be seen in Figure \ref{fig:gain_vs_hv_rt_ct}. The slope of the curves follows a simple power law. For the same HV, the gain at CT is lower than at RT. Being the gain 10$^7$ at RT, it decreases by 76\% due to cryogenic conditions. An increase of 170\,V is required on this PMT to compensate the gain loss at CT.

The gain is also measured at a temperature closer to the one in ProtoDUNE-DP, which will be 94 K, taking into account the pressure at the bottom of the LAr cryostat. To achieve a higher temperature, an over-pressure of 1 bar is applied to the vessel, which is the maximum allowed by the set-up. While the temperature with LN$_2$ at atmospheric pressure is 77\,K, during these tests 83\,K are reached. Gain variations are found (see Figure~\ref{fig:gain_vs_hv_rt_ct}) compatible with expectations~\cite{clean1}. For the same PMT, a  10$^7$ gain at RT decreases 66\% at 83\,K. This means that an increase of 130\,V in the HV is required at RT to obtain the same gain. In total, 4 PMTs were measured at 83 K, and the gain decreases on average 60$\pm$9\%.

\subsection{Linearity}
\label{sec5.3}

\begin {figure}[ht]
\includegraphics[width=0.45\textwidth]{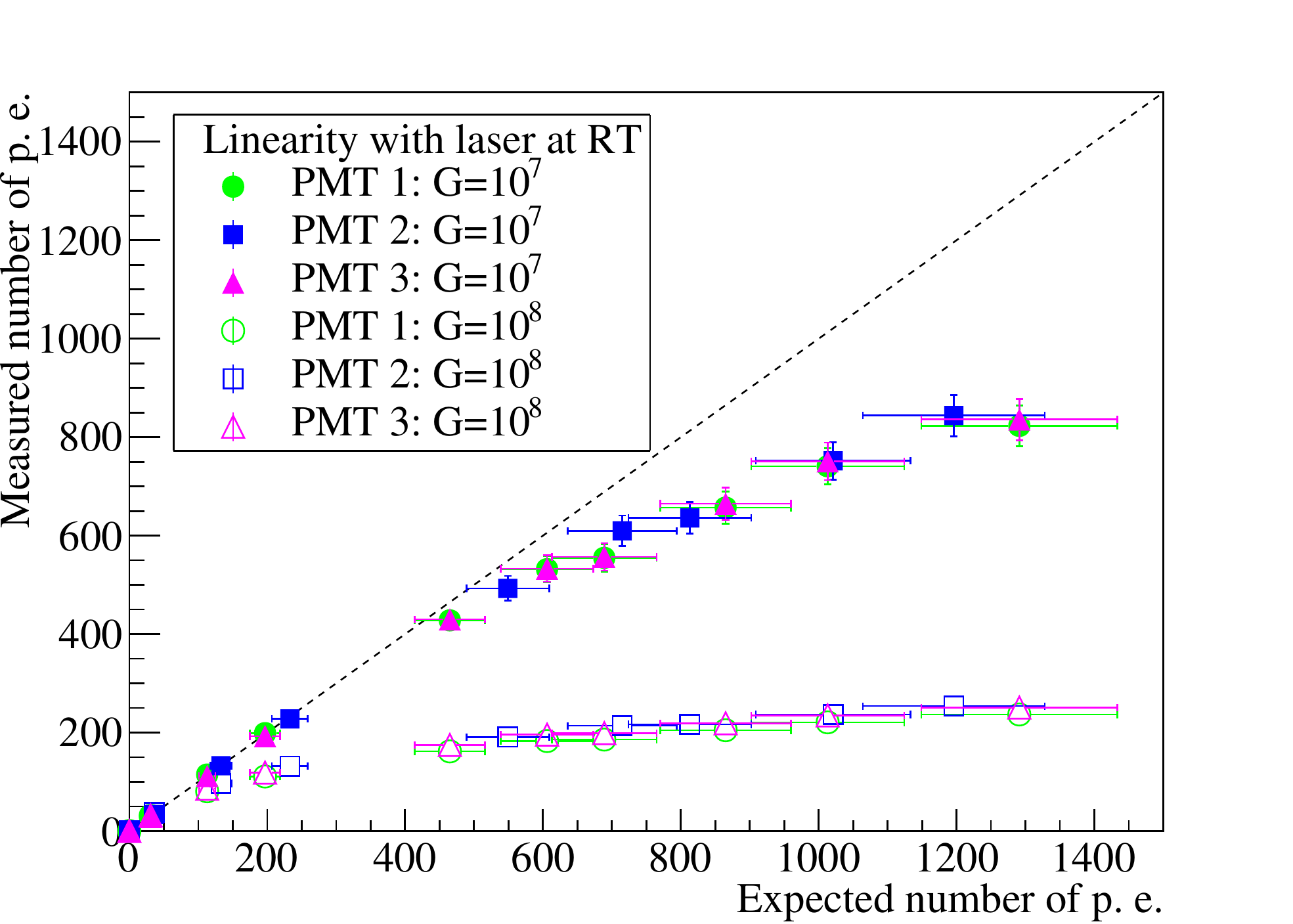}
\centering \caption{Measured vs expected number of p.e illuminating three PMTs at different gains at RT with the laser.}
\label{fig:lin1}
\end {figure}

The response of several PMTs as a function of the light intensity is measured at RT and CT. The response at RT for $\sim$10$^7$ and $\sim$10$^8$ gains illuminating a PMT with the laser is shown in Figure~\ref{fig:lin1}.  For gains $<$10$^6$ the response has been observed to be linear in the tested range (up to $\sim$1000 p.e.). For gains close to 10$^7$, the PMT remains linear up to 200 p.e., while for larger gains ($>$10$^8$) the PMT response deviates from linearity with only 75 p.e., and a saturation regime is reached. Same response is observed for all tested PMTs, as an example the perfect agreement of three PMTs is shown in Figure~\ref{fig:lin1}. The uncertainty in the measured number of p.e. is 4\% and comes from the gain variation obtained fitting spectra at different light levels. On the other hand, the error in the expected number of p.e. is 9\% and corresponds to error propagation of the uncertainties in the gain estimation and in the measurement of the transmission of the filters.

\begin {figure}[ht]
\subfigure[]{\includegraphics[width=0.49\textwidth]{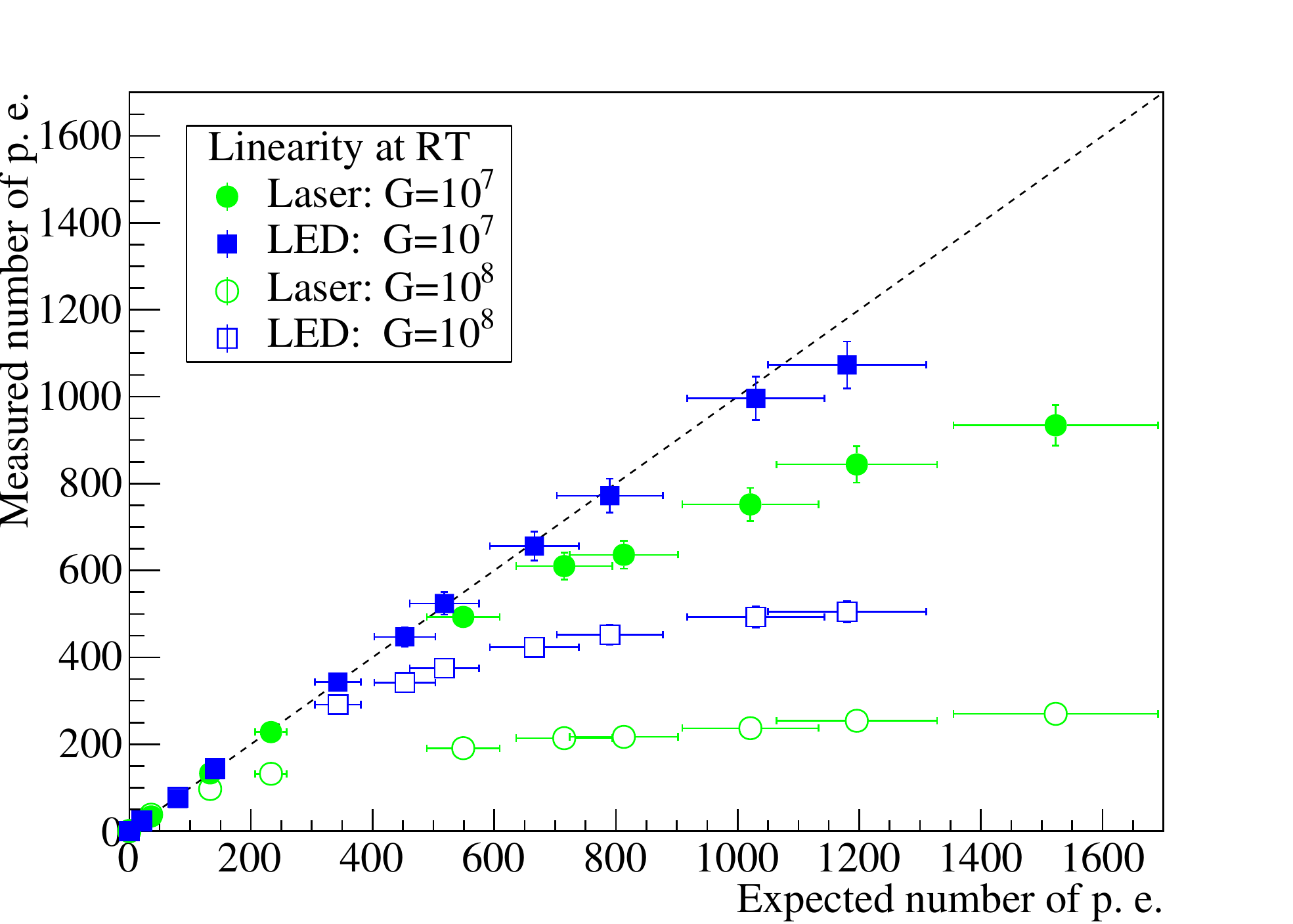}\label{fig:lin2a}}
\subfigure[]{\includegraphics[width=0.49\textwidth]{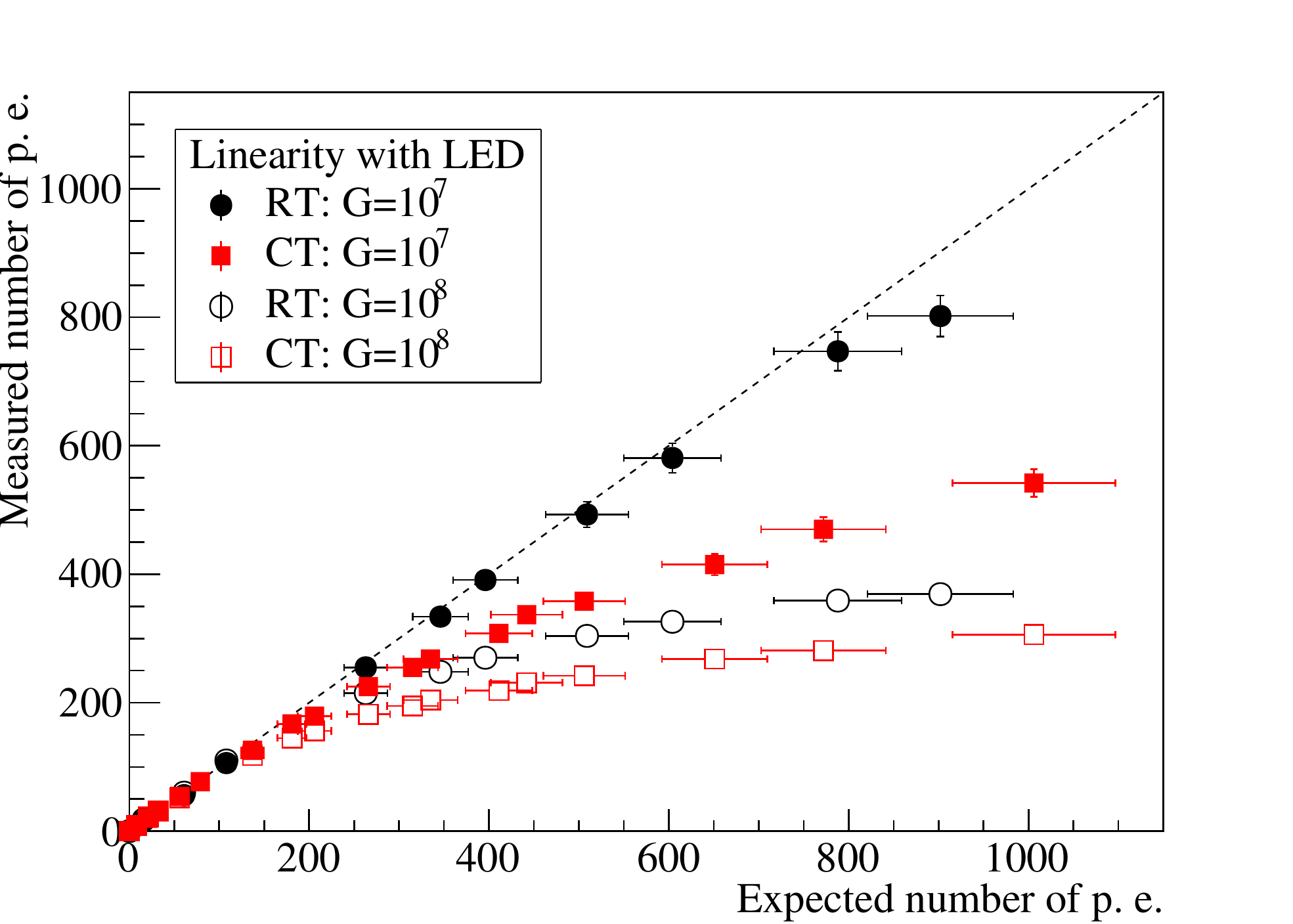}\label{fig:lin2b}}
\centering \caption{Comparison of the measured vs expected number of p.e. illuminating the PMT with the LED and the laser at RT (a) and comparison of RT and CT for the LED (b).}
\label{fig:lin2}
\end {figure}

The PMTs are also illuminated with the LED, and the results are compared to the laser ones in Fig.~\ref{fig:lin2a}. The linearity range increases when the light source is wider, showing a clear dependency with the pulse shape in the PMT saturation. At CT, the PMT response is just slightly worse than at RT when the PMT is illuminated by the laser. For instance, for a 10$^8$ gain and 500 p.e., the measured number of p.e. is 60\% and 70\% lower than expected at RT and CT, respectively. However, as can be seen in Fig.~\ref{fig:lin2b}, the linear range when the PMT is illuminated by the LED at CT is shorter than at RT, and similar to the laser at CT. Therefore, despite the very good linearity at RT, the PMT response is saturated at 300 p.e. for a gain of 10$^7$ at CT.

\subsection{Light rate}
\label{sec5.4}

The PMT response is studied for pulsed frequencies from 100\,Hz to 10\,MHz using the laser and the LED as light sources and different light intensities (from 10  to 50\,p.e.). 

\begin {figure}[ht]
\subfigure[]{\includegraphics[width=0.49\textwidth]{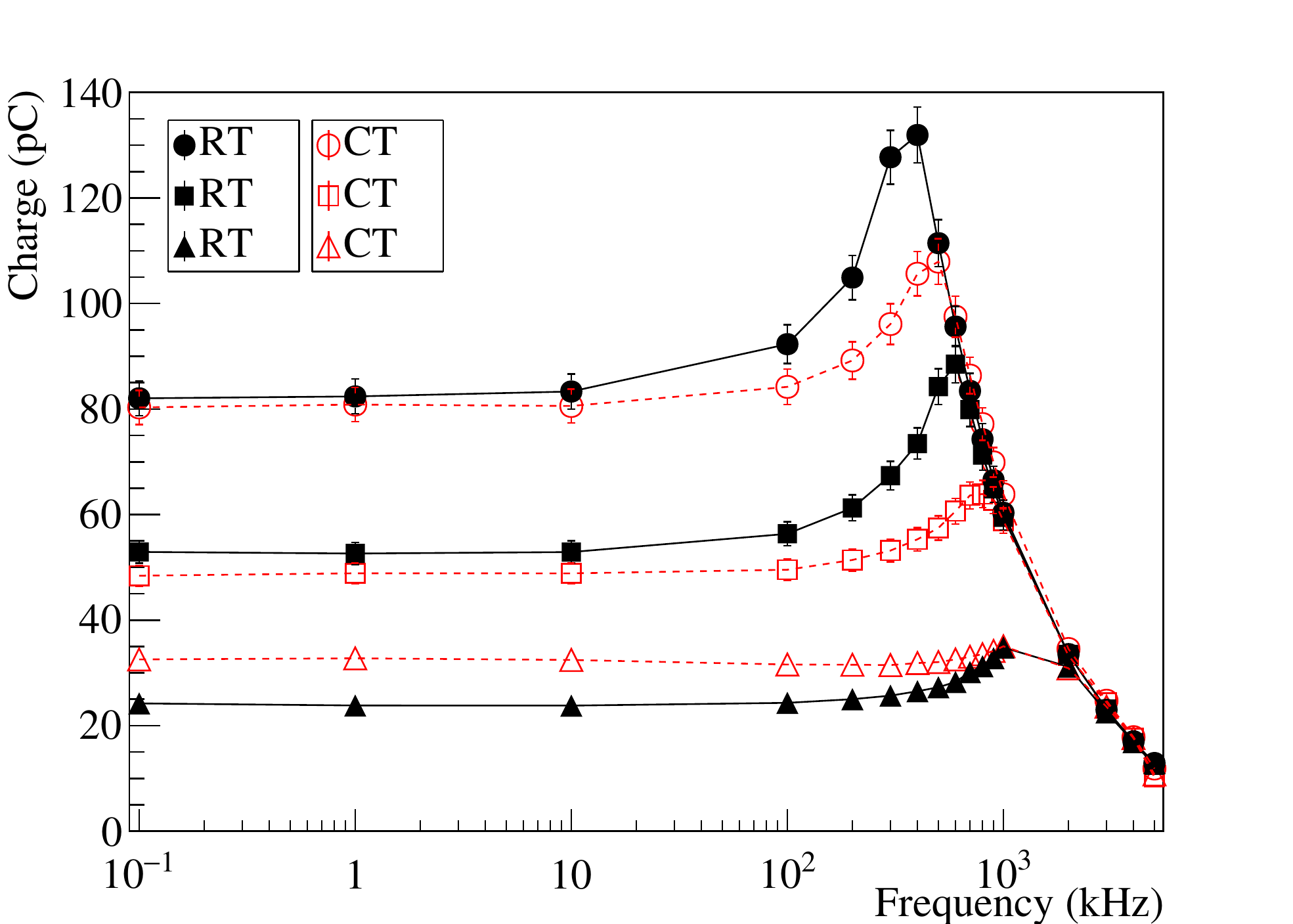}\label{fig:freqa}}
\subfigure[]{\includegraphics[width=0.49\textwidth]{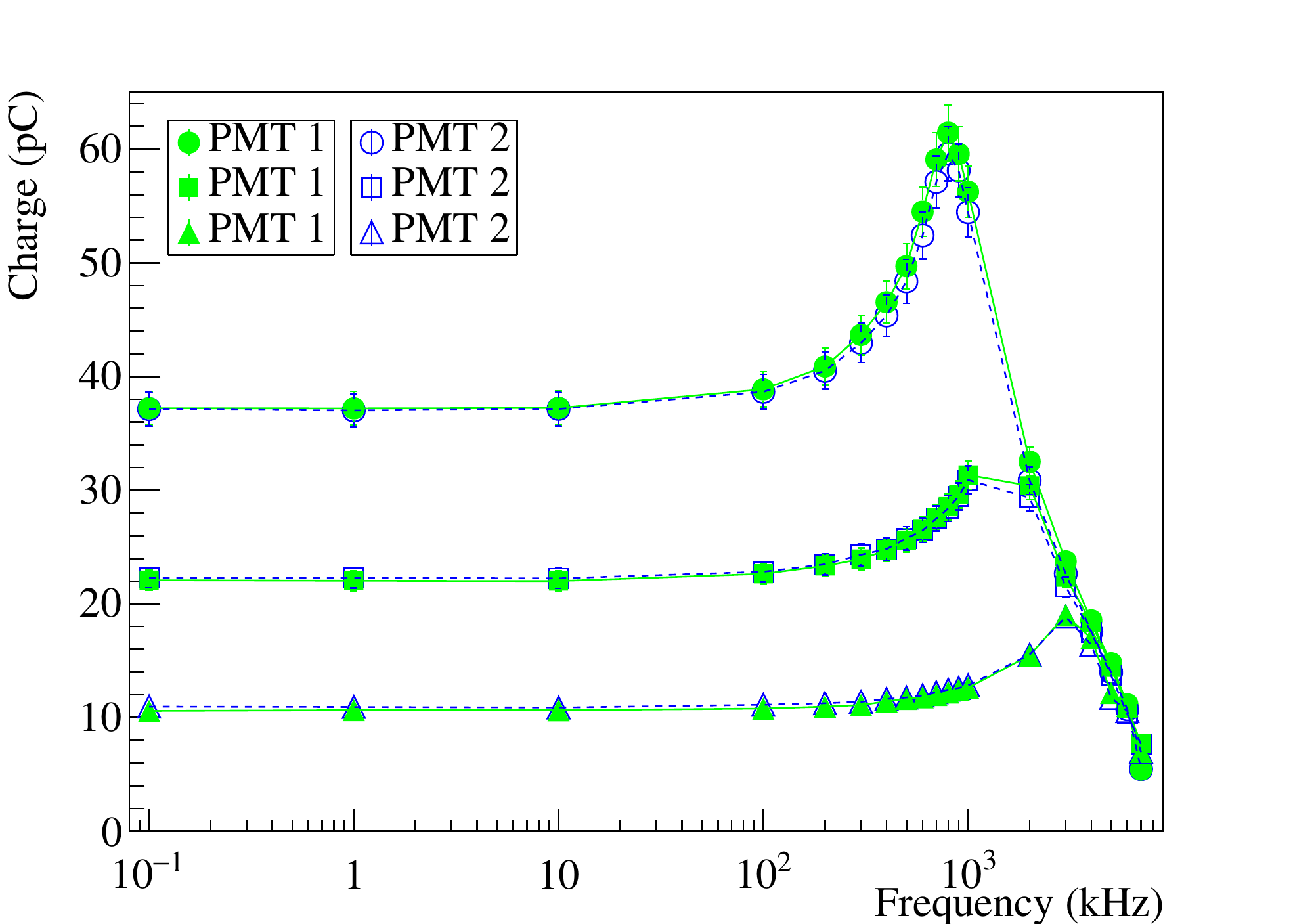}\label{fig:freqb}}
\centering \caption{(a) PMT LED frequency response at RT and CT for different amounts of
light. (b) Two PMTs frequency response comparison for laser at RT. The vertical error bars are 4\% and are given by the variation of the charge obtained when repeating the measurements.}
\label{fig:freq}
\end {figure}

At RT, see Figure \ref{fig:freqa}, the three regions explained in section \ref{sec4.4} are observed: the PMT response is flat until a given frequency ($>$10\,kHz) which depends on the charge; then, the over-linearity effect is observed; and the PMT saturation ($>$500\,kHz). At CT, the PMT response is expected to be the same as at RT, see Figure \ref{fig:freqa}, as the saturation curve depends only on the base design.  The expected PMT gain reduction, as the average output current increases (see section \ref{sec4.2}), compensates and reduces the over-linearity region. The same result is observed for different PMTs, see Figure \ref{fig:freqb}.

\begin {figure}[ht]
\includegraphics[width=0.49\textwidth]{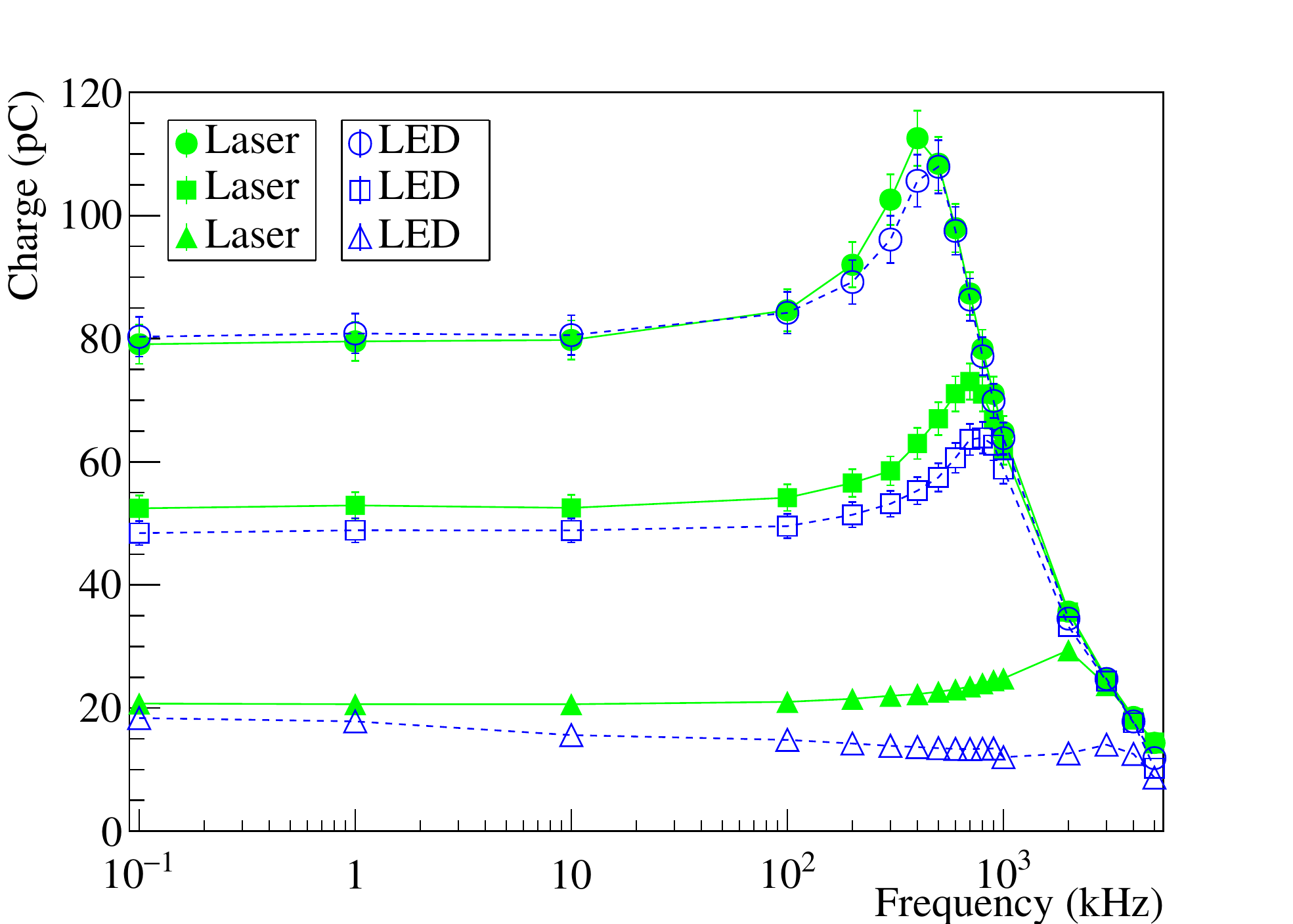}
\centering \caption{Charge vs frequency for LED and laser with similar charge at CT. The vertical error bars are 4\% and are given by the variation of the charge obtained when repeating the measurements.}
\label{fig:freq3}
\end {figure}

The frequency sweep is also done using the LED, and the results, shown in Figure \ref{fig:freq3}, suggest that the response is very similar, but the over-linearity starts at slightly higher frequencies when the LED is used. As the LED light pulses are wider, the PMT output peak current is smaller (for the same charge) moving the over-linearity effect to higher frequencies. For low intensity signals, as expected for the S2 light in ProtoDUNE-DP, the PMT response is linear until $\sim$1\,MHz, far enough for our purpose.

\subsection{PMT base validation}
\label{sec5.5}

In order to validate the design of the PMT base, the PB and NB are compared at RT and CT in terms of gain, dark rate, and linearity. As expected, the PMT gain with both configurations is similar.  In order to compare DC rate vs voltage dependence, both bases are tested on the same PMT and on the same darkness conditions. The results show that the behavior of both bases is similar at voltages up to 1600\,V, and at higher HV the DC rate increases more on the NB reaching a rate around 50\% higher than the PB at 1900\,V. The base with negative power supply shows higher dark rate than the positive one because the photocathode is at high voltage and spurious pulses could appear due to current leakage through the glass or due to electro-luminescence in the glass.

\begin {figure}[ht]
\subfigure[]{\includegraphics[width=0.49\textwidth]{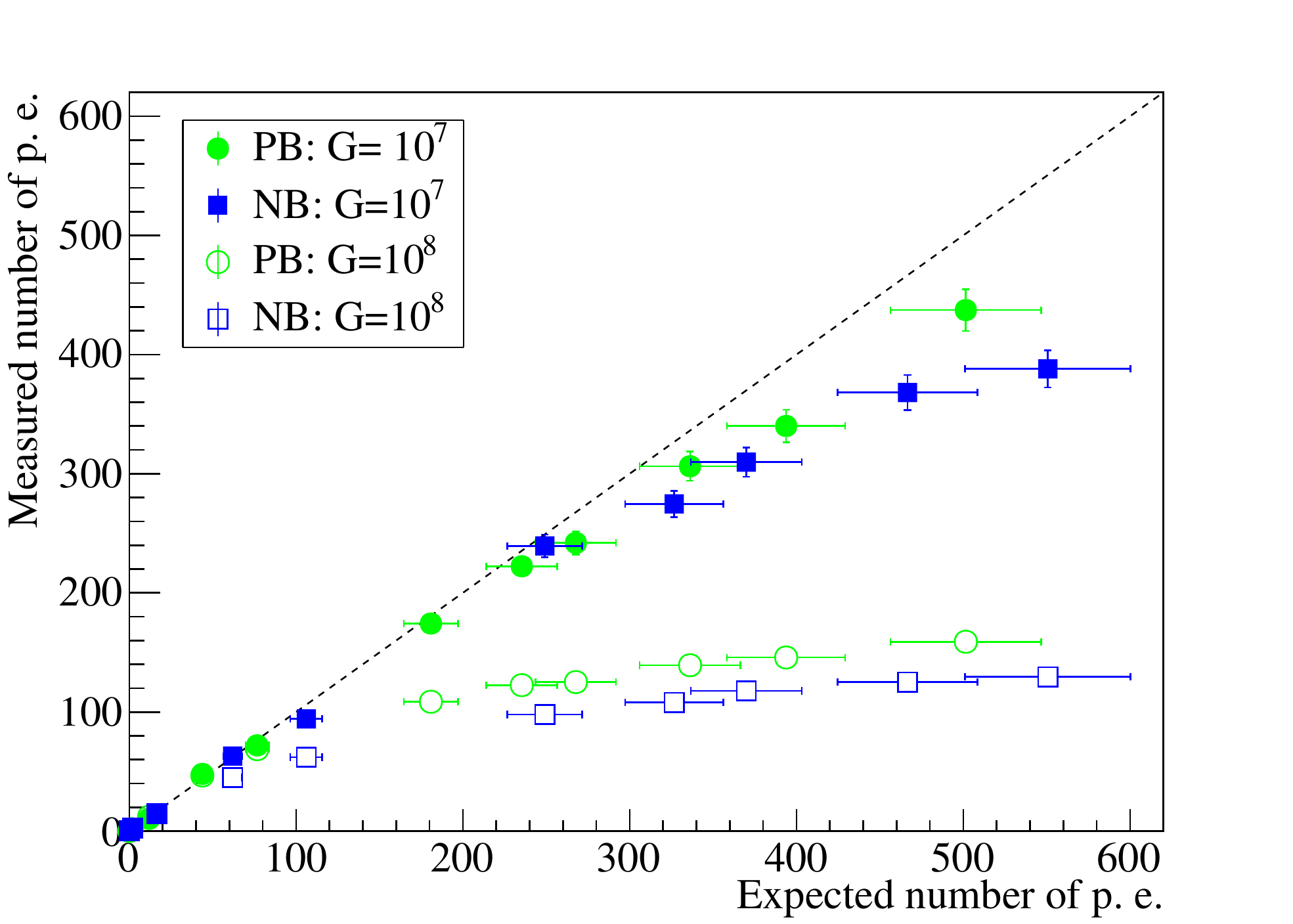}\label{fig:PBNBa}}
\subfigure[]{\includegraphics[width=0.49\textwidth]{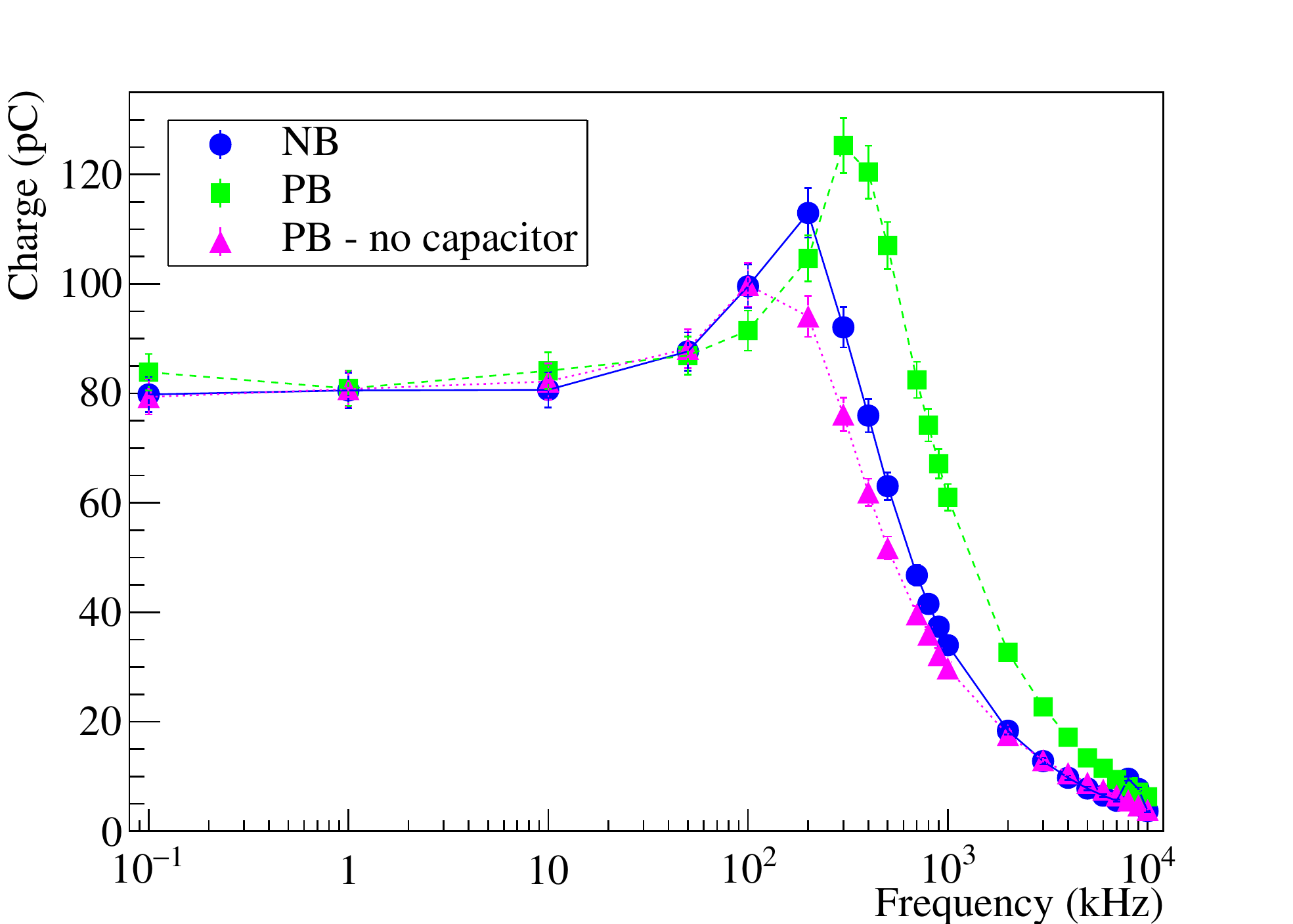}\label{fig:PBNBb}}
\centering \caption{(a) Measured vs expected number of p.e. for the NB and PB at RT with the laser. (b) Response vs light rate for NB, PB and PB without filtering capacitor at RT with the laser.}
\label{fig:PBNB}
\end {figure}

For the linearity response, the positive base shows slightly better results, see Figure~\ref{fig:PBNBa}. On the positive bias base, the power supply filter capacitor is closer to the anode which increases the charge reservoir for the PMT output increasing slightly the linearity range.


For the study of the response with the light rate, the tests are done following the procedure described in section~\ref{sec4.4}. The positive circuit shows also better results, as shown in Figure~\ref{fig:PBNBb}. The difference between the two bases is also due to the different position of the filtering capacitor on the base. To verify it, a PB without this capacitor was also tested and its behavior was slightly worse than with the NB, see Figure~\ref{fig:PBNBb}, making clear that the presence of this capacitor close to the anode improves the linearity of the PMT. Increasing the capacity of this capacitor did not improve the response any further at the tested light levels. These tests confirm the better performance of the PB.

\section{ProtoDUNE-DP PMT characterization}
\label{sec6}

Last, all the PMTs to be installed at ProtoDUNE-DP are characterized. Measurements with 40 (36 + 4 spares) PMTs are taken to verify the successful functioning and create a database to be used during the detector commissioning and operation. DC and gain as a function of HV at RT and CT are measured. Also, typical waveforms for all the PMTs at 10$^7$ gain are recorded.

\subsection{Dark Current}
\label{sec6.1}

\begin{figure}[ht!]
\centering
\subfigure[]{\includegraphics[width=0.49\textwidth]{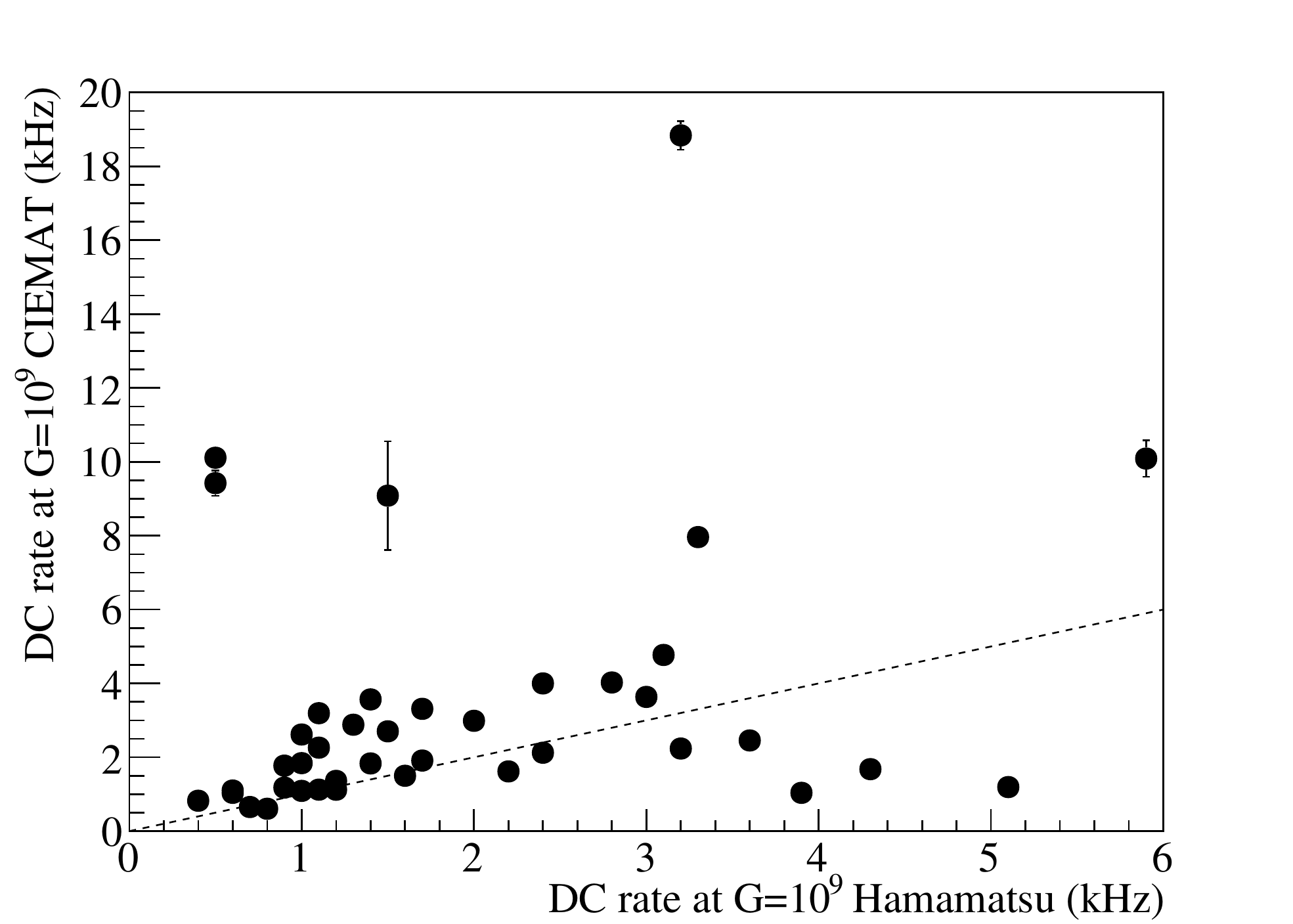}\label{fig:dca}}
\subfigure[]{\includegraphics[width=0.49\textwidth]{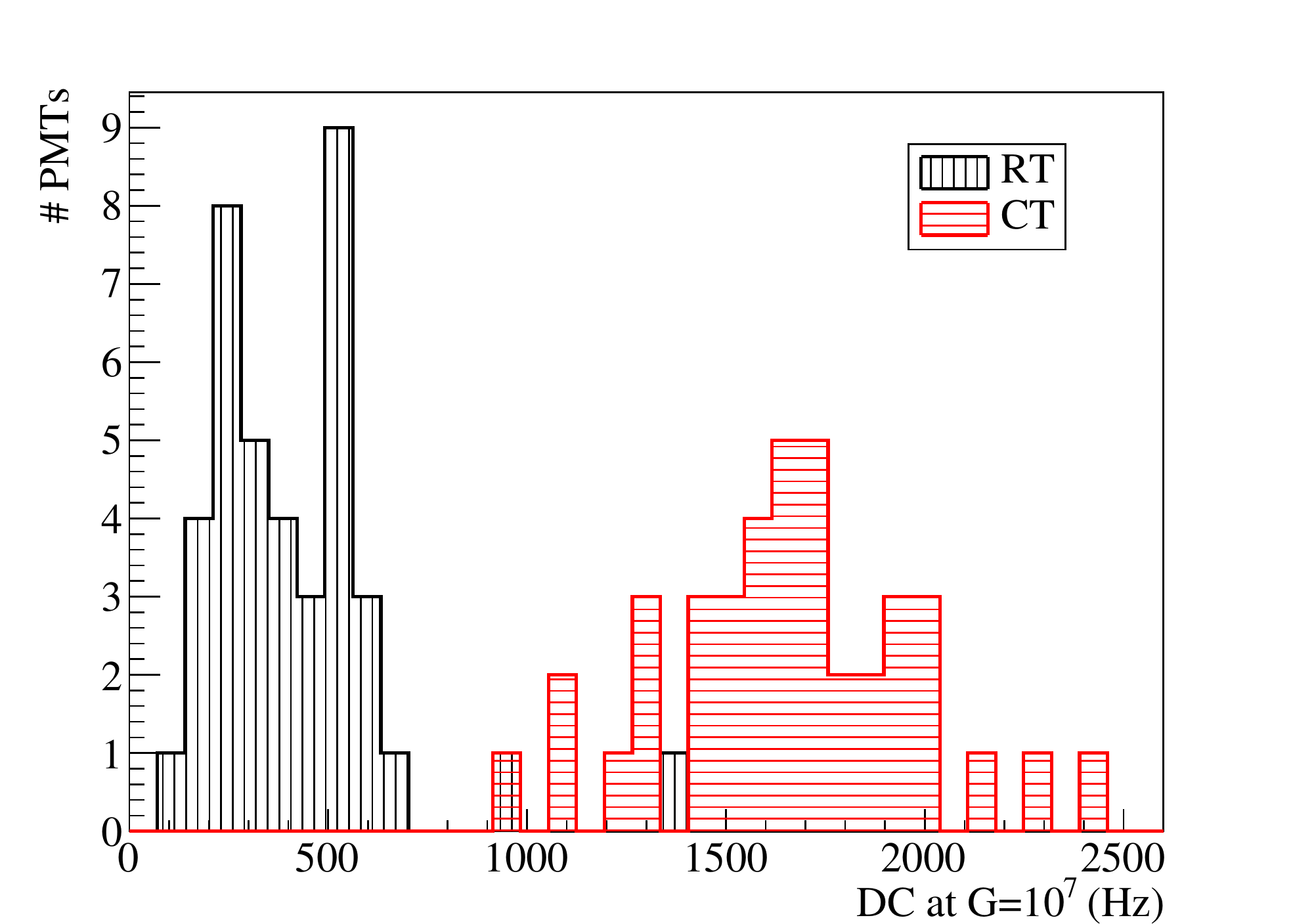}\label{fig:dcb}}
\caption{(a) DC measured at CIEMAT vs Hamamatsu for $10^9$ gain for the 40 PMTs. DC is measured every 10\,s during 5\,min, and the error bars represent the rms of the average value. The dotted line represents the identity line. (b) DC histograms for 10$^7$ gain at RT and CT measured at CIEMAT for the 40 PMTs. At RT, DC rate is on average 0.4\,$\pm$\,0.2\,kHz, and at CT, 1.7\,$\pm$\,0.3\,kHz. }
\label{fig:dc_ciemat_hamamatsu}
\end{figure}

The DC rate is measured for the 40 PMTs at several HVs as explained in section \ref{sec4.1}. In particular, the DC rate at $10^9$ gain is measured to compare with the value given by the manufacturer. Figure~\ref{fig:dca} shows the correlation between the rates measured at CIEMAT (at the HV specified by the manufacturer) and the ones provided by Hamamatsu at RT. In general, similar results are obtained. However, two of them had to be replaced, as one had no signal and another one showed a DC rate of around 100 kHz (almost 60 times the expected rate from Hamamatsu). It is worth noting that the defective PMTs are not shown in the plot, but the replacements are shown instead.

On average, as shown in Figure~\ref{fig:dcb}, DC rate is 0.4\,$\pm$\,0.2\,kHz when the PMTs operate at $\sim$10$^7$ gain at RT, and DC rate is always below 1.4\,kHz. However, at CT, DC rate increases up to 1.7\,$\pm$\,0.3\,kHz, and some PMTs reach up to 2.5\,kHz. No correlation between DC at RT and CT is observed. The PMTs with very unstable DC rates in time (i.e. large error bars in the Fig.~\ref{fig:dca}) were inspected by Hamamatsu and although no defects were found, they can be designated as spares. The PMTs with the lowest DC rate are selected and the rest is left as spare.

\subsection{Gain}
\label{sec6.2}

\begin{figure}[ht!]
\centering
\includegraphics[width=0.49\textwidth]{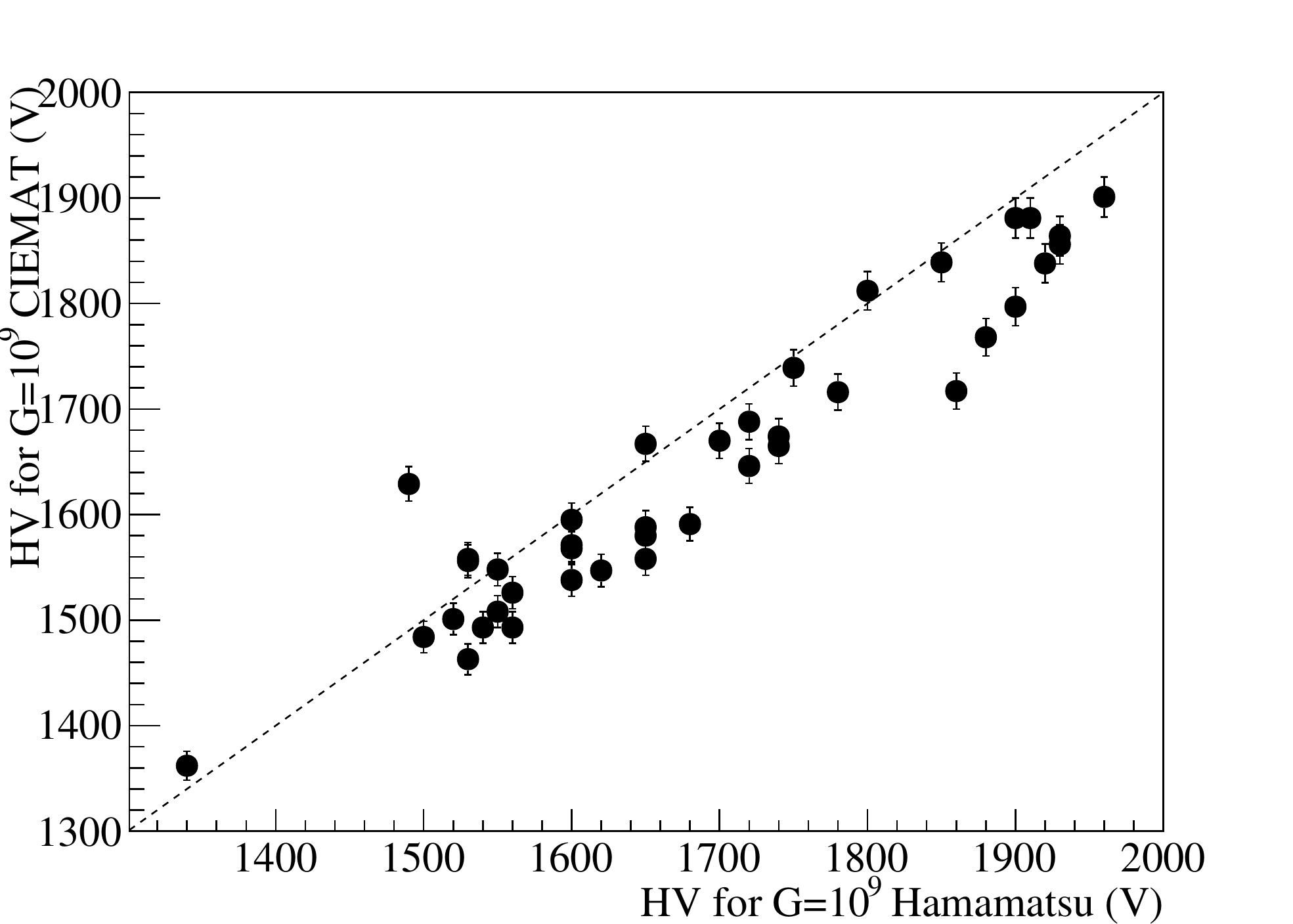}
\caption{HV applied at CIEMAT compared to the HV provided by Hamamatsu for $10^9$ gain for the 40 PMTs. The dotted line represents the identity line.}
\label{fig:hv_ciemat_hamamatsu}
\end{figure}

\begin{figure}[ht!]
\centering
\subfigure[]{\includegraphics[width=0.49\textwidth]{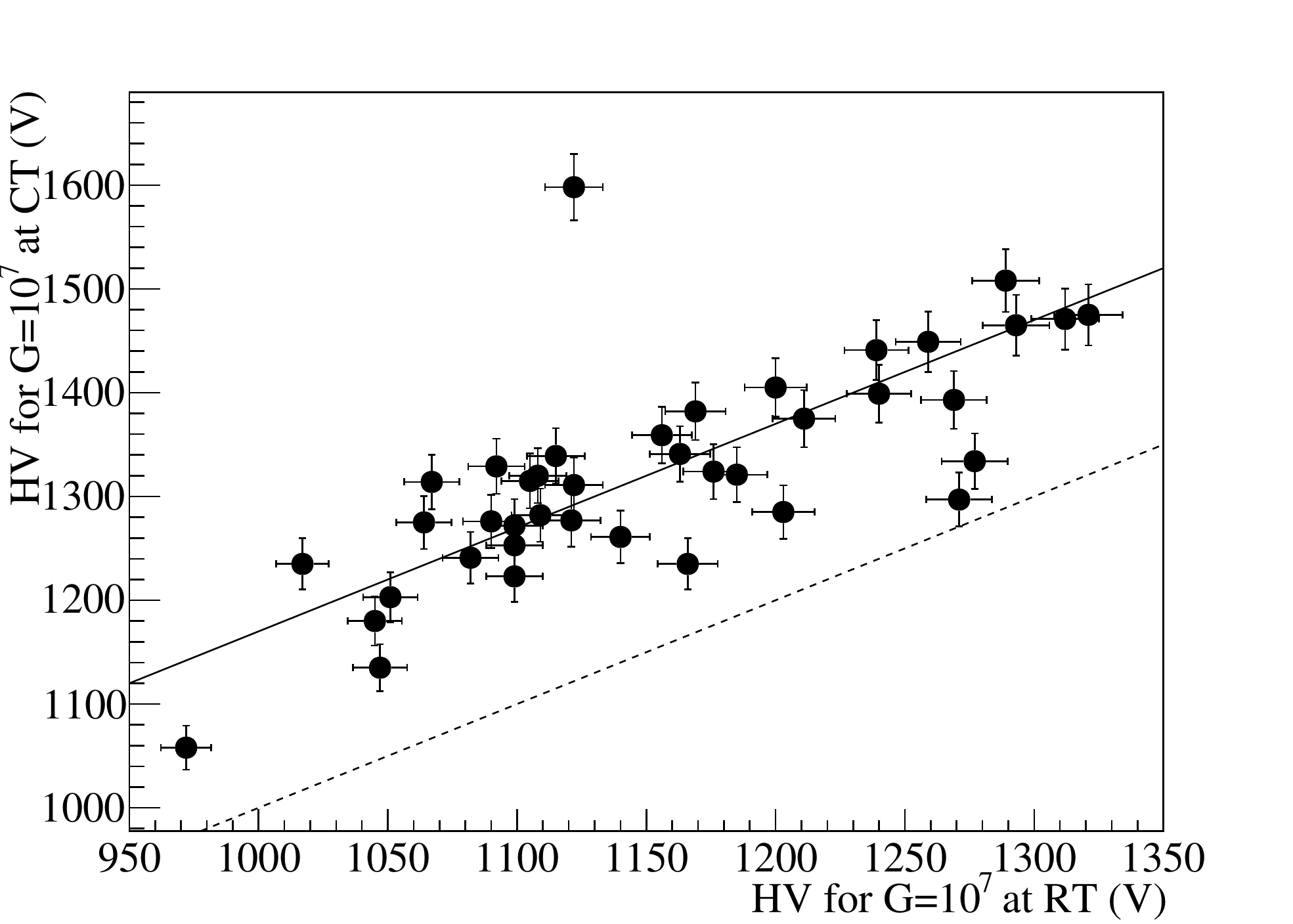}\label{fig:hv_ciemat_hamamatsub}}
\subfigure[]{\includegraphics[width=0.49\textwidth]{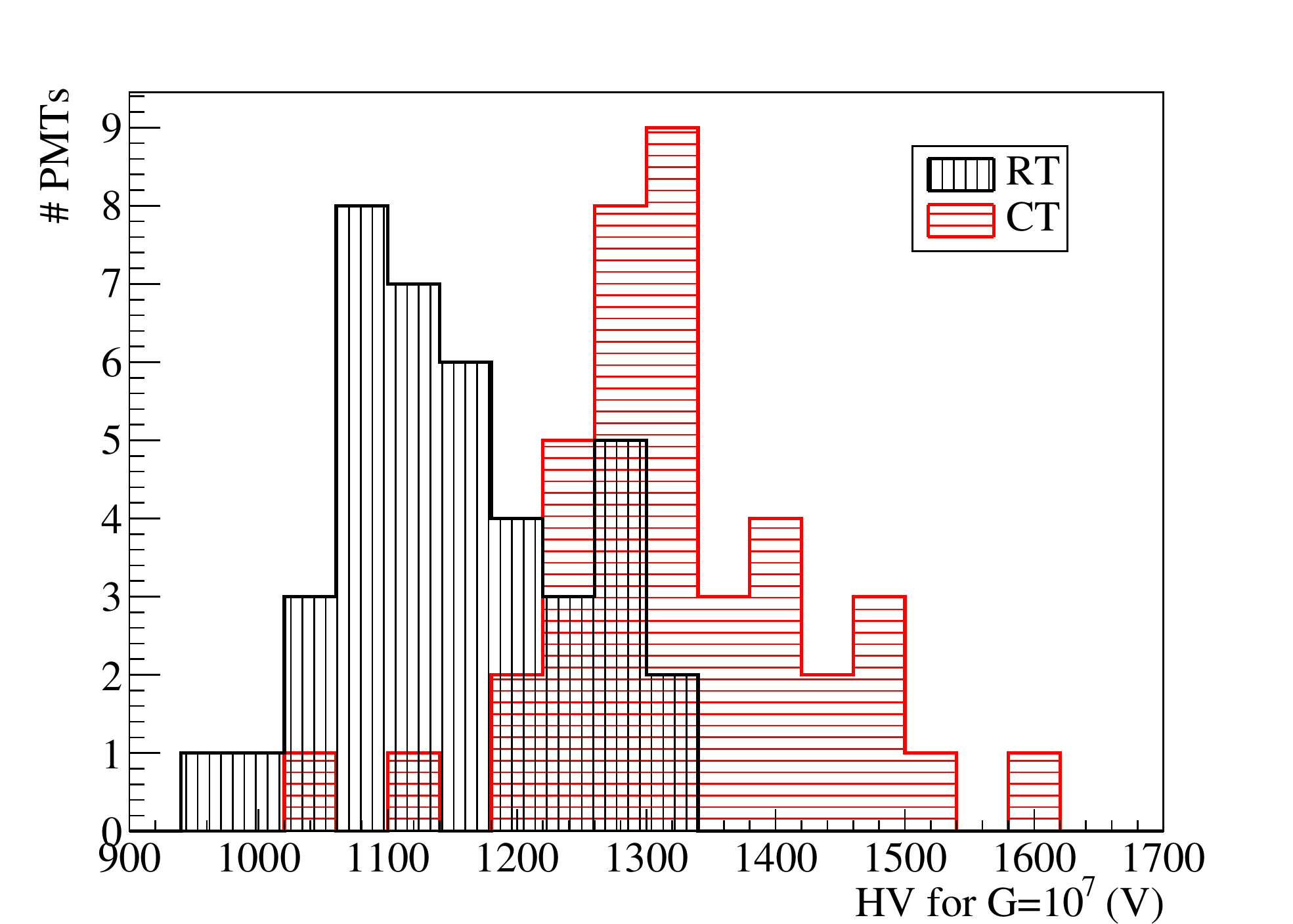}\label{fig:hv_ciemat_hamamatsuc}}
\caption{(a) HV at RT vs CT for $10^7$ gain measured at CIEMAT for the 40 PMTs.  The dotted line represents the identity line and the solid line the identity line shifted 170\,V. (b) Histogram of HV required to obtain a  $10^7$ gain at RT and CT. The average HV is 1154$\pm$87\,V at RT, and 1324$\pm$103\,V at CT.}
\label{fig:hv_ciemat}
\end{figure}

The gain results from the characterization of the 40 PMTs are presented here. Figure~\ref{fig:hv_ciemat_hamamatsu} shows the correlation between the HV determined at CIEMAT for a $10^9$ gain, and the HV required according to manufacturer specifications (both values at RT). The CIEMAT HV is extrapolated from the gain-voltage curve of each PMT. A good correlation is observed with only 2.3\% deviation possibly due to differences in the setup or in the gain determination method. In Figure~\ref{fig:hv_ciemat}, the HV required for a $10^7$ gain at RT and CT is presented. Higher HV, 170$\pm$72 V on average, needs to be applied at CT to reach the same gain, which is equivalent to 71$\pm$14\% gain drop at CT.

\section*{Conclusions}
\addcontentsline{toc}{section}{\protect\numberline{}Conclusions}%

The ProtoDUNE-DP experiment aims to build and operate a LAr TPC detector at CERN to fully demonstrate the dual-phase technology at large scale for DUNE, the next generation long-baseline neutrino experiment. The photon detection system will add precise timing capabilities, and will be formed by 8-inch cryogenic photomultipliers from Hamamatsu positioned at the bottom of the detector. The validation measurements of the PMT model (R5912-20Mod) and the base design at cryogenic temperature are described. In addition, the 40 PMTs to be used in ProtoDUNE-DP are characterized to ensure their performance according to the specifications and valuable results to be used by other experiments are obtained.

It is observed that the dark current rate at CT is higher than at RT due to non-thermal contributions, i.e. at a gain of 10$^7$ the DC rate is on average  to 1.7$\pm$0.3\,kHz at CT while at RT is 0.4$\pm$0.2\,kHz. This effect has been observed previously  in different PMTs models, being proportional to the photocathode area \cite{meyer2}. From RT to CT, the gain decreases, and a gain of  $10^7$ at RT, drops by 71$\pm$14\% when the PMT operates at 77\,K , and 60$\pm$9\% at 83\,K. 

At CT, a fatigue effect is observed as the PMT output current increases, either, due to high gain, high light intensity or high light rate. The gain recovery time depends on the PMT excitation (output current) and varies from PMT to PMT. The largest effect is observed, as expected, when the excitation comes from high frequency signals of $\sim$MHz.

The linearity of the PMT response with the incident light depends on the PMT gain and the width of the incident light pulses. For fast signals (<1ns pulse) the PMT remains linear up to at least 1000 p.e. for a 10$^6$ gain, but loses linearity at 75 p.e. in the case of 10$^8$ gain. However, when the same amount of light is distributed on a wider (40 ns) pulse, the linearity region of the PMT increases. A slightly worse behavior is observed at CT for the laser, but an earlier saturation is observed for the LED in comparison to RT.

PMT response with the light rate is flat until >10\,kHz for less than 100\,p.e.. At CT, the PMT response is expected to be the same as at RT, as the saturation curve depends only on the base design, but gain reduction compensates and reduces the over-linearity region. 

For these reasons, characterization tests at CT before installing the PMTs are required, and a dedicated photon calibration system is recommended to monitor the PMT gain during the experiment data taking period. It is concluded that these PMTs are validated and will be used in ProtoDUNE-DP.

\acknowledgments
This project has received funding from the European Union Horizon~2020 Research and Innovation programme under Grant Agreement no.~654168 and from the Spanish Ministerio de Econom\'ia y Competitividad (SEIDI-MINECO) under Grants no.~FPA2016-77347-C2-1-P, FPA2016-77347-C2-2-P, MdM-2015-0509, and SEV-2016-0588.

\bibliographystyle{JHEP}
\bibliography{PMTpaper_revision}

\end{document}